# IS THE MOYAL EQUATION FOR THE WIGNER FUNCTION A QUANTUM ANALOGUE OF THE LIOUVILLE EQUATION?

**E.E. Perepelkin**[a,b,d*], **B.I. Sadovnikov**[a], **N.G. Inozemtseva**[b,c], **E.V. Burlakov**[b], **P.V. Afonin**[a]

[a] *Faculty of Physics, Lomonosov Moscow State University, Moscow, 119991 Russia*
[b] *Moscow Technical University of Communications and Informatics, Moscow, 123423 Russia*
[c] *Dubna State University, Moscow region, Dubna,141980 Russia*
[d] *Joint Institute for Nuclear Research, Moscow region, Dubna,141980 Russia*
*Corresponding author: pevgeny@jinr.ru

**Abstract**

The Moyal equation describes the evolution of the Wigner function of a quantum system in the phase space. The right-hand side of the equation contains an infinite series with coefficients proportional to powers of the Planck constant. There is an interpretation of the Moyal equation as a quantum analogue of the classical Liouville equation. Indeed, if one uses the notion of the classical passage to the limit as the Planck constant tends to zero, then formally the right-hand side of the Moyal equation tends to zero. As a result, the Moyal equation becomes the classical Liouville equation for the distribution function.

In this paper, we show that the right side of the Moyal equation does not explicitly depend on the Planck constant, and all terms of the series can make a significant contribution. The transition between the classical and quantum descriptions is related not to the Planck constant, but to the spatial scale.

For a model quantum system with a potential in the form of a «quadratic funnel», an exact 3D solution of the Schrödinger equation is found and the corresponding Wigner function is constructed in the paper. Using trajectory analysis in the phase space, based on the representation of the right-hand side of the Moyal equation, it is shown that on the spatial microscale there is an infinite number of «trajectories» of the particle motion (thereby the concept of a trajectory is indefinite), and when passing to the macroscale, all «trajectories» concentrate around the classical trajectory.

**Introduction**

The classical and quantum descriptions of physical systems significantly differ. The concept of a trajectory, which is central to classical physics, is absent in the quantum description of the microworld. The introduction by Wigner [1-2] of the phase space for quantum systems, despite the Heisenberg uncertainty principle, made it possible to analyze the processes in the microworld in terms of classical physics. The Wigner function $W(\vec{r},\vec{p},t)$ of the quasi-probability density of a quantum system in the phase space has areas of negative values [3-6], and its evolution is described by the Moyal equation [7]:

$$\frac{\partial}{\partial t}W + \frac{\vec{p}}{m}\cdot\nabla_r W - \nabla_r U \cdot \nabla_p W = \sum_{l=1}^{+\infty}\frac{(-1)^l (\hbar/2)^{2l}}{(2l+1)!}U\left(\overleftarrow{\nabla}_r \cdot \overrightarrow{\nabla}_p\right)^{2l+1}W, \qquad (\text{i.1})$$

where $U$ is the potential in the Schrödinger equation. The Moyal equation (i.1) can be derived from the von Neumann equation for the matrix of density $\hat{\rho}$ and the representation of the Wigner function

$$W(\vec{r},\vec{p},t) = \frac{1}{(2\pi\hbar)^3} \int_{\mathbb{R}^3} \left\langle \vec{r} - \frac{\vec{s}}{2} \left| \hat{\rho} \right| \vec{r} + \frac{\vec{s}}{2} \right\rangle e^{-i\frac{\vec{p}\cdot\vec{s}}{\hbar}} d^3s. \tag{i.2}$$

The right-hand side of equation (i.1) contains a series with coefficients $\hbar^{2l}$, where $\hbar$ is the Planck constant. If for equation (i.1) one formally performs the «classical passage to the limit» at $\hbar \to 0$, then the Moyal equation (i.1) becomes the classical Liouville equation for distribution function $f(\vec{r},\vec{p},t)$ with a zero right-hand side. As a result, equation (i.1) looks like an extension of the Liouville equation for the case of quantum systems.

When deriving the Moyal equation, it is assumed that potential $U$ is an analytic function, that is, it can be expanded in a power series. In the case of a polynomial potential of the form:

$$U_N(x,y,z,t) = \sum_{k_1,k_2,k_3=0}^{N} \left( a_{k_1} x^{k_1} + b_{k_2} y^{k_2} + c_{k_3} z^{k_3} \right), \tag{i.3}$$

where $a_{k_1}, b_{k_2}, c_{k_3}$ are known coefficients, there was constructed in [8] the Vlasov-Moyal approximation of mean kinematical value $\langle \dot{\vec{v}} \rangle_{1,2}$ of the quantum system described by the Wigner function (i.2):

$$\langle \dot{v}_\mu \rangle_{1,2} = \sum_{l=0}^{+\infty} \frac{(-1)^{l+1} (\hbar/2)^{2l}}{m^{2l+1}(2l+1)!} \frac{\partial^{2l+1} U}{\partial x_\mu^{2l+1}} \frac{1}{f^{1,2}} \frac{\partial^{2l} f^{1,2}}{\partial v_\mu^{2l}}, \tag{i.4}$$

where $f^{1,2}(\vec{r},\vec{v},t) = m^3 W(\vec{r},\vec{p},t)$. Substituting approximation (i.4) into the second Vlasov equation [9] for function $f^{1,2}$

$$\frac{\partial}{\partial t} f^{1,2} + \vec{v} \cdot \nabla_r f^{1,2} + \mathrm{div}_v \left[ f^{1,2} \langle \dot{\vec{v}} \rangle_{1,2} \right] = 0, \tag{i.5}$$

transforms equation (i.5) into the Moyal equation (i.1). Thus, the Moyal equation is a particular case of the second Vlasov equation (i.5).

Note that when multiplying equation (i.5) by the component of velocity $v_\mu$ and then integrating it over the entire velocity space, the equation of motion in the hydrodynamic approximation is obtained [9]:

$$\frac{d}{dt} \langle v_\mu \rangle_1 = \left( \frac{\partial}{\partial t} + \langle v_\lambda \rangle_1 \frac{\partial}{\partial x_\lambda} \right) \langle v_\mu \rangle_1 = -\frac{1}{f^1} \frac{\partial P_{\mu\lambda}}{\partial x_\lambda} + \langle \dot{v}_\mu \rangle_1, \tag{i.6}$$

$$P_{\mu\lambda} = \int_{(\infty)} f^{1,2} \left( v_\mu - \langle v_\mu \rangle_1 \right) \left( v_\lambda - \langle v_\lambda \rangle_1 \right) d^3 v, \tag{i.7}$$

where $P_{\mu\lambda}$ is the pressure tensor, and mean kinematical values $\langle v_\mu \rangle$ and $\langle \dot{v}_\mu \rangle$ satisfy the relations:

$$f^1\langle v_\mu\rangle_1 = \int_{(\infty)} f^{1,2} v_\mu d^3v, \quad f^1\langle \dot v_\mu\rangle_1 = \int_{(\infty)} f^{1,2} \langle \dot v_\mu\rangle_{1,2} d^3v, \quad f^1 = \int_{(\infty)} f^{1,2} d^3v. \qquad (i.8)$$

Function $f^1$ is a solution to the first Vlasov equation (i.9), which is obtained by integrating equation (i.5) over the space of velocities (i.8):

$$\frac{\partial}{\partial t} f^1 + \mathrm{div}_r\left[ f^1 \langle \vec v\rangle \right] = 0. \qquad (i.9)$$

For arbitrary analytic potential $U$, constructing an approximation of kinematical value $\langle \dot{\vec v}\rangle_{1,2}$ is not a trivial task. Such an uncertainty in the form of field $\langle \dot{\vec v}\rangle_{1,2}$ arises from the ambiguous transition from quasi-probability density function (the Wigner function) $f^{1,2} = m^3 W$ to the vector field $\langle \dot{\vec v}\rangle_{1,2}$, which cannot be definitely unique. The quantum formalism works with a scalar value in the form of wave function $\Psi$, a probability density $|\Psi|^2$ or a quasi-density of probabilities $W$. Classical physics, the Vlasov kinematical equation (i.5) contains mean vector field of accelerations $\langle \dot{\vec v}\rangle_{1,2}$. In quantum formalism, there is no concept of a particle trajectory, which requires the presence of vector value $\langle \dot{\vec v}\rangle_{1,2}$ inherent in classical physics. It is believed that the Schrödinger equation and the wave function carry all the information about the evolution of a quantum system [10]. An attempt to interpret the behavior of a quantum system in terms of classical physics requires additional information. The absence of this information generates uncertainty in the form of field $\langle \dot{\vec v}\rangle_{1,2}$.

Note that the first summand in approximation (i.4) corresponds to the classical equation of motion $m\langle \dot{\vec v}\rangle_{1,2} = -\nabla_r U + ...$ (i.6). The subsequent terms of series (i.4) have coefficients $\hbar^{2l}$, which should disappear in the formal passage to the limit to classical mechanics $\hbar \to 0$.

The aim of this work is to study the correctness of passage to the limit $\hbar \to 0$ in equation (i.1) and options for constructing approximations of type (i.4). The paper shows that the transition from the quantum to the classical description in equation (i.1) is generally associated not with classical limit $\hbar \to 0$, but with the spatial scale on which the physical system is considered.

The work has the following structure. §1 considers the mathematical formulation of the problem of restoring vector field $\langle \dot{\vec v}\rangle_{1,2}$ from its divergence and curl (the Helmholtz problem). The equations of quantum mechanics contain information only about the divergence of the current density field $\vec J^2_{1,2} = f^{1,2}\langle \dot{\vec v}\rangle_{1,2}$. The lack of information about the current curl $\vec J^2_{1,2}$ leads to the uncertainty of the acceleration field and, as a result, to different trajectories of motion. In § 2 various options are considered for constructing approximations of type (i.4) for the general case of analytic potential $U$. It is shown that there is an infinite set of such options, which has a cardinality — a continuum. It turns out that the terms of series (i.1) do not explicitly contain factor $\hbar^{2l}$ and cannot be «omitted» in classical limit $\hbar \to 0$ as small values. Moreover, at small spatial scales («microscales»), the terms of series (i.1) can make a significant contribution to field $\langle \dot{\vec v}\rangle_{1,2}$. On large spatial scales («macroscales»), the terms in series (i.1) can be negligibly small, and the Moyal equation turns into the Liouville equation. §3 presents numerical calculations of a model quantum system with potential in the form of a «quadratic funnel»:

$U(\rho, z) \sim \left( \rho^2 + z^2 - 4\frac{\sigma_r^4}{\rho^2} \right)$, where $(\rho, z)$ is a cylindrical coordinate system. The exact solution of the 3D Schrödinger equation with potential $U(\rho, z)$ is found and the Wigner function is constructed. In a particular case, the Wigner function can be expressed in terms of a complete elliptic integral of the first kind. A numerical-analytical scheme for finding field $\langle \dot{\vec{v}} \rangle_{1,2}$ from the Wigner function and the potential $U(\rho, z)$ is constructed. The conducted trajectory analysis shows that on microscales ($r < \sigma_r$) there is an infinite number of different trajectories of particle motion. The presence of an infinite set of trajectories introduces uncertainty in coordinate and momentum. On macroscales ($r > \sigma_r$), the set of trajectories «concentrates» around the only classical trajectory described by the equation of motion $m \langle \dot{\vec{v}} \rangle_{1,2} = -\nabla_r U$, for which the coordinate and momentum are determined with acceptable accuracy. The Appendix section contains the proofs of the theorems and intermediate mathematical calculations.

## §1 Construction of the approximation

Based on the Moyal equation (i.1) and the second Vlasov equation (i.5), the relation must be satisfied:

$$\operatorname{div}_v \vec{J}_{1,2}^2 = \operatorname{div}_v \left[ f^{1,2} \langle \dot{\vec{v}} \rangle_{1,2} \right] = \sum_{l=0}^{+\infty} \frac{(-1)^{l+1} (\hbar/2)^{2l}}{m^{2l+1} (2l+1)!} U \left( \overleftarrow{\nabla}_r \cdot \vec{\nabla}_v \right)^{2l+1} f^{1,2}. \quad (1.1)$$

If function $f^{1,2}$ satisfies the Vlasov condition at infinity [9], then field $\vec{J}_{1,2}^2$ has a fast decay ($1/v^{2+\varepsilon}$, $\varepsilon > 0$). According to the Helmholtz theorem [11], sufficiently smooth vector field $\vec{J}_{1,2}^2$ can be represented as the sum of a potential and a vortex field:

$$\vec{J}_{1,2}^2 = \vec{J}_{1,2}^{2(p)} + \vec{J}_{1,2}^{2(s)} = -\nabla_v \Pi^{1,2} + \operatorname{curl}_v \vec{\Upsilon}_{1,2}, \quad (1.2)$$

where indices «$p$» and «$s$» denote the potential and vortex components of field $\vec{J}_{1,2}^2$, respectively.

Expression (1.1) contains information only about the potential component of field $\vec{J}_{1,2}^2$, that is $\operatorname{div}_v \vec{J}_{1,2}^2 = -\Delta_v \Pi^{1,2}$. Using expression (1.1) it is possible to find potential $\Pi$ with an accuracy up to a harmonic function

$$\Pi^{1,2}(\vec{r}, \vec{v}, t) = \frac{1}{4\pi} \int_{(\infty)} \frac{1}{|\vec{v} - \vec{v}'|} \operatorname{div}_v \vec{J}_{1,2}^2 d^3 v'. \quad (1.3)$$

**Theorem 1** *Let condition (1.1) be satisfied and series (1.1) admit term-by-term integration, then the expression for the scalar potential (1.3) has the form:*

$$\Pi^{1,2} = \sum_{l=0}^{+\infty} \frac{(-1)^{l+1} (\hbar/2)^{2l}}{m^{2l+1} (2l+1)!} U \left( \overleftarrow{\nabla}_r \cdot \vec{\nabla}_v \right)^{2l+1} \vartheta^{1,2}, \quad (1.4)$$

*where*

$$\vartheta^{1,2}(\vec{r},\vec{v},t) \overset{\text{det}}{=} \frac{1}{4\pi} \int_{(\infty)} \frac{f^{1,2}(\vec{r},\vec{v}',t)}{|\vec{v}-\vec{v}'|} d^3v'. \quad (1.5)$$

*wherein*

$$\Delta_v \vartheta^{1,2} = -f^{1,2}. \quad (1.6)$$

The proof of the theorem is given in Appendix A.

**Remark 1**

From the point of view of potential theory, function $\vartheta$ (1.5) defines the scalar potential created by sources distributed in the space of velocities with density $f^{1,2}(\vec{r},\vec{v},t)$. Equation (1.6) is the Poisson equation:

$$\text{div}_v \, \vec{D}^2_{1,2} = f^{1,2}. \quad (1.7)$$

According to the Helmholtz theorem, field $\vec{D}^2_{1,2}$ can be represented as

$$\vec{D}^2_{1,2} = -\partial_1 \vec{A}^2_{1,2} - \nabla_v \vartheta, \quad (1.8)$$

where $\partial_1 = \partial_0 + \vec{v}\cdot\nabla_r$

$$\text{div}_v \, \vec{A}^2_{1,2} = \text{div}_r \, \vec{A}^2_{1,2} = 0. \quad (1.9)$$

The second condition $\text{div}_r \, \vec{A}^2_{1,2} = 0$ in (1.8) is additional to the coordinate dependence of vector potential $\vec{A}^2_{1,2}$. If there is field $\vec{A}^2_{1,2}$ satisfying conditions (1.9), then representation (1.8) is correct, that is $\text{div}_v(\partial_1 \vec{A}^2_{1,2}) = 0$, and equation (1.7) is also satisfied. Indeed,

$$\text{div}_v(\partial_1 \vec{A}_{1,2}) = \partial_0 \, \text{div}_v \, \vec{A}_{1,2} + \text{div}_v\left[(\vec{v}\cdot\nabla_r)\vec{A}_{1,2}\right] = \partial_0 \, \text{div}_v \, \vec{A}_{1,2} + \text{div}_r \, \vec{A}_{1,2} + (\vec{v}\cdot\nabla_r)\text{div}_v \, \vec{A}_{1,2},$$

$$\text{div}_v(\partial_1 \vec{A}^2_{1,2}) = \partial_1 \, \text{div}_v \, \vec{A}^2_{1,2} + \text{div}_r \, \vec{A}^2_{1,2}, \quad (1.10)$$

where $\partial_0 = \partial/\partial t$ and it is taken into account that

$$\frac{\partial}{\partial v_\alpha}\left(v_\beta \frac{\partial}{\partial x_\beta} A^2_\alpha\right) = \delta_{\alpha\beta}\frac{\partial}{\partial x_\beta}A^2_\alpha + v_\beta \frac{\partial}{\partial x_\beta}\frac{\partial}{\partial v_\alpha}A^2_\alpha = \frac{\partial}{\partial x_\alpha}A^2_\alpha + v_\beta \frac{\partial}{\partial x_\beta}\frac{\partial}{\partial v_\alpha}A^2_\alpha.$$

Expression (1.10) proves the validity of representation (1.8) for equation (1.7). Since function $f^{1,2}(\vec{r},\vec{v},t)$ satisfies the second Vlasov equation, then, according to [12], we obtain equations for fields $\vec{D}^1_{1,2}$ and $\vec{D}^2_{1,2}$:

$$f^{1,2} = \text{div}_r \, \vec{D}^1_{1,2} + \text{div}_v \, \vec{D}^2_{1,2}, \quad (1.11)$$

where

$$\text{div}_r \, \vec{D}^1_{1,2} = 0. \quad (1.12)$$

Equation (1.11) is an extension of original equation (1.7). Substituting (1.11) into the second Vlasov equation, we obtain

$$\partial_1 f^{1,2} + \mathrm{div}_v \left[ f^{1,2} \langle \dot{\vec{v}} \rangle_{1,2} \right] = 0,$$

$$\mathrm{div}_r \partial_0 \vec{D}^1_{1,2} + \mathrm{div}_v \partial_0 \vec{D}^2_{1,2} + \vec{v} \nabla_r \, \mathrm{div}_r \vec{D}^1_{1,2} + \vec{v} \nabla_r \, \mathrm{div}_v \vec{D}^2_{1,2} + \mathrm{div}_v \vec{J}^2_{1,2} = 0. \quad (1.13)$$

We take into account that

$$\vec{v} \nabla_r \, \mathrm{div}_r \vec{D}^1_{1,2} \mapsto v_\beta \frac{\partial}{\partial x_\beta} \frac{\partial D^1_\alpha}{\partial x_\alpha} = \frac{\partial}{\partial x_\alpha} v_\beta \frac{\partial D^1_\alpha}{\partial x_\beta} \mapsto \mathrm{div}_r \left[ (\vec{v} \cdot \nabla_r) \vec{D}^1_{1,2} \right], \quad (1.14)$$

$$\vec{v} \nabla_r \, \mathrm{div}_v \vec{D}^2_{1,2} \mapsto v_\beta \frac{\partial}{\partial x_\beta} \frac{\partial}{\partial v_\alpha} D^2_\alpha = \frac{\partial}{\partial v_\alpha} \left( v_\beta \frac{\partial}{\partial x_\beta} D^2_\alpha \right) - \delta_{\alpha\beta} \frac{\partial}{\partial x_\beta} D^2_\alpha = \frac{\partial}{\partial v_\alpha} \left( v_\beta \frac{\partial}{\partial x_\beta} D^2_\alpha \right) - \frac{\partial}{\partial x_\alpha} D^2_\alpha,$$

$$\vec{v} \nabla_r \, \mathrm{div}_v \vec{D}^2_{1,2} = \mathrm{div}_v \left[ (\vec{v} \cdot \nabla_r) \vec{D}^2_{1,2} \right] - \mathrm{div}_r \vec{D}^2_{1,2}. \quad (1.15)$$

Substitution of expressions (1.14) and (1.15) into equation (1.13) leads to the relation:

$$\mathrm{div}_r \left[ \partial_0 \vec{D}^1_{1,2} + (\vec{v} \cdot \nabla_r) \vec{D}^1_{1,2} + \vec{J}^1_{1,2} \right] + \mathrm{div}_v \left[ \partial_0 \vec{D}^2_{1,2} + (\vec{v} \cdot \nabla_r) \vec{D}^2_{1,2} + \vec{J}^2_{1,2} \right] = 0, \quad (1.16)$$

where $\vec{J}^1_{1,2} \stackrel{\mathrm{det}}{=} -\vec{D}^2_{1,2}$. Relation (1.16) can be satisfied under the conditions [12]:

$$\partial_1 \vec{D}^1_{1,2} + \vec{J}^1_{1,2} = \mathrm{curl}_r \vec{H}^1_{1,2}, \quad (1.17)$$

$$\partial_1 \vec{D}^2_{1,2} + \vec{J}^2_{1,2} = \mathrm{curl}_v \vec{H}^2_{1,2}, \quad (1.18)$$

where $\vec{H}^1_{1,2}$ and $\vec{H}^2_{1,2}$ are some fields. In [12], the Maxwell equations of the second rank for the dispersive chain of equations of quantum mechanics are given, which coincide with (1.11), (1.17) and (1.18) under the condition $\vec{H}^1_{1,2} = 0$ and $\mathrm{div}_r \vec{D}^1_{1,2}$ different from zero. In case $\vec{H}^1_{1,2} \neq 0$ one can get the equation for the field $\vec{H}^1_{1,2}$. It follows from a comparison of representation (1.8) and equation (1.17) that

$$\vec{D}^2_{1,2} = \partial_1 \vec{D}^1_{1,2} - \mathrm{curl}_r \vec{H}^1_{1,2} = -\partial_1 \vec{A}^2_{1,2} - \nabla_v \vartheta,$$

that is

$$\vec{A}^2_{1,2} = -\vec{D}^1_{1,2}, \quad \mathrm{div}_r \vec{D}^1_{1,2} = -\mathrm{div}_r \vec{A}^2_{1,2} = 0, \quad (1.19)$$

$$\mathrm{curl}_r \vec{H}^1_{1,2} = \nabla_v \vartheta, \quad (1.20)$$

where condition (1.9) is taken into account. Let us take $\mathrm{div}_v$ from equation (1.20), we obtain

$$\mathrm{div}_v \mathrm{curl}_r \vec{H}^1_{1,2} = \Delta_v \vartheta = -f^{1,2},$$

$$\varepsilon_{\beta\alpha\lambda} \frac{\partial^2 H^1_\lambda}{\partial x_\alpha \partial v_\beta} = -f^{1,2}, \quad (1.21)$$

where $\varepsilon_{\beta\alpha\lambda}$ is the Levi-Civita symbol. If equation (1.21) is solvable, then its solution determines $\vec{H}_{1,2}^1$ field up to some vortex function in the space of velocities.

If series (1.4) admits term-by-term differentiation, then it follows from Theorem 1 that potential component $\vec{J}_{1,2}^{2(p)}$ of the probability current density (1.2) has the form

$$J_\mu^{2(p)} = -\frac{\partial}{\partial v_\mu}\Pi^{1,2} = -\sum_{l=0}^{+\infty}\frac{(-1)^{l+1}(\hbar/2)^{2l}}{m^{2l+1}(2l+1)!}U\left(\tilde{\vec{\nabla}}_r,\vec{\nabla}_v\right)^{2l+1}\frac{\partial \mathcal{G}^{1,2}}{\partial v_\mu}. \quad (1.22)$$

**Theorem 2** *Let current field $\vec{J}_{1,2}^2$ (1.2) be represented in the form (1.2)-(1.6), and series (1.22) admit term-by-term integration, then the Vlasov condition [9] of zero flow over the velocity space is satisfied:*

$$\int_{\Sigma_\infty} \vec{J}_{1,2}^2 \cdot d\vec{\sigma}_v = 0, \quad (1.23)$$

*where $\Sigma_\infty$ is an infinitely distant surface.*

The proof of Theorem 2 is given in Appendix A.

Note that when obtaining the chain of Vlasov equations, it is assumed that the condition of rapid decay to zero of distribution function $f^{1,2}$ on infinitely distant surface $\Sigma_\infty$ is satisfied, which guarantees the fulfillment of equality (1.23). In the case under consideration (1.1), in the right-hand side, $f^{1,2}$ is initially replaced by the Wigner function, which does not have to satisfy the Vlasov condition. Therefore, the fulfillment of Theorem 2 guarantees a fast decay of the Wigner function at infinity.

**§2 Multiple-valuedness and «quantum» indeterminacy**

Representation (1.22) makes it possible to construct potential component $J_\mu^{2(p)}$ of the current density $\vec{J}_{1,2}^2$. Vortex component $\vec{\Upsilon}_{1,2}$ (1.2) remains unknown, since equation (1.1) contains information only about scalar field sources. From a mathematical point of view, there are two problems. The first problem is the lack of information about vortex component $\vec{\Upsilon}_{1,2}$ of field (1.2). Relation (1.1) contains information only about the scalar sources of field $\vec{J}_{1,2}^2$. If we assume that there is a method or an equation from which vortex field $\vec{\Upsilon}_{1,2}$ can be found, then the second problem arises, connected with the indeterminacy of the solution of the Helmholtz problem (1.2). For example, an arbitrary harmonic function $P^{1,2}$ can be added to potential $\Pi^{1,2}$:

$$\Pi^{1,2} \mapsto \Pi^{1,2} + P^{1,2}, \text{ where } \Delta_v P^{1,2} = 0. \quad (2.1)$$

Adding field $\nabla_v P^{1,2}$ to expansion (1.2) will not change the expressions for $\text{div}_v \vec{J}_{1,2}^2$ and $\text{curl}_v \vec{J}_{1,2}^2$. The statement of the problem of restoring field $\vec{J}_{1,2}^2$ with respect to div and curl has three types:

1. an inner problem (in a bounded domain $\Omega_i$ with smooth boundary $\Gamma_i$ and a condition on it);
2. an external problem (in unbounded domain with a notch $\Omega_e = \mathbb{R}^3 \setminus \Omega_{cut}$ and a boundary condition at $\Gamma_e$);
3. a problem for the entire space $\mathbb{R}^3$ (it is assumed that the Vlasov condition is satisfied at infinity).

All three statements, have a unique solution provided that a solution exists. In formulations 1 and 2, the uniqueness of the solution is determined by the boundary conditions at boundaries $\Gamma_{e,i}$, which determine the choice of the function $P^{1,2}$. A Dirichlet, Neumann, or mixed type condition can be specified at boundary $\Gamma_{e,i}$.

Note that potential $P^{1,2}$ is a solution of the elliptic type equation (2.1). Thus, the values of function $P^{1,2}$ inside domain $\Omega_{e,i}$ are determined by the values at each point at boundary $\Gamma_{e,i}$. If at least one point at boundary $\Gamma_{e,i}$ changes the value of potential $P^{1,2}$ (or its normal derivative), then field $\nabla_v P^{1,2}$ inside domain $\Omega_{e,i}$ will be different. ***Such a mathematical connection is similar to the nonlocality property of a physical system.***

According to the maximum principle for harmonic functions, potential $P^{1,2}$ takes on its maximum and minimum values at boundary $\Gamma_{e,i}$. If it is hypothetically possible to localize a physical system in a sufficiently large domain, at the boundary of which potential $P^{1,2}$ value tends to zero (the Vlasov condition), then, according to the Liouville theorem, function $P^{1,2}$ will tend to a constant value. From a formal point of view, the uncertainty caused by field $\nabla_v P^{1,2}$ will disappear. The described situation corresponds to the third formulation of the Helmholtz problem for the entire space.

**Remark 2**

Thus, when transiting from a quantum description to a classical one, an indeterminacy arises in the construction of field $\vec{J}^2_{1,2}$ due to the lack of information about vortex component $\vec{\Upsilon}_{1,2}$ and the boundary conditions for the first and second statements (which are associated with the nonlocality of the system). Despite the possible different types of approximations of the kinematical mean $\langle \dot{\vec{v}} \rangle_{1,2}$, the solution of the Vlasov/Moyal equation will remain unchanged, since it depends only on the scalar sources of the field (1.1), which are uniquely determined.

In addition to the Helmholtz method described in §1, we can consider the construction of field $J^{2(p)}_\mu$ according to expression (1.1) from «physics» considerations. First, we obtain a simplified version for the representation of field $\vec{J}^2_{1,2}$. Expression (1.22) requires knowledge of the potential $\vartheta^{1,2}$, which is sought by integrating over the entire velocity space (1.5). Let us consider approximation $\vec{J}^2_{1,2}$ of the form:

**Definition** *Approximation of the mean field of accelerations* $\langle \dot{v}_\mu \rangle_{1,2}$ *of the form:*

$$f^{1,2} \langle \dot{v}_\mu \rangle_{1,2} \stackrel{\text{det}}{=} \sum_{l=0}^{+\infty} \frac{(-1)^{l+1}(\hbar/2)^{2l}}{m^{2l+1}(2l+1)!} \frac{\partial U}{\partial x_\mu} \left( \vec{\nabla}_r \cdot \vec{\nabla}_v \right)^{2l} f^{1,2}, \qquad (2.2)$$

*will be called the Vlasov-Moyal approximation for arbitrary analytic potential $U(\vec{r},t)$.*

**Theorem 3** *If series (2.2) admits term-by-term integration and differentiation, then the substitution of approximation (2.2) into the second Vlasov equation (i.5) gives the Moyal equation (i.1) for the Wigner function $f^{1,2}(\vec{r},\vec{v},t) = m^3 W(\vec{r},\vec{p},t)$, wherein*

$$\langle \dot{v}_\mu \rangle_1 = \frac{1}{f^1} \int\limits_{(\infty)} f^{1,2} \langle \dot{v}_\mu \rangle_{1,2} d^3v = -\frac{1}{m}\frac{\partial U}{\partial x_\mu}. \quad (2.3)$$

The proof of Theorem 3 is given in Appendix A.

**Remark 3**

Approximation (2.2) has a number of differences from approximation (1.22). First, from a mathematical point of view, expression (2.2) is simpler than (1.22), since it does not require finding additional potential $\vartheta^{1,2}$ (1.5). Secondly, (2.2) may contain the vortex components of field $\vec{J}_{1,2}^2$:

$$\text{curl}_v \vec{J}_{1,2}^2 = \sum_{l=0}^{+\infty} \frac{(-1)^l (\hbar/2)^{2l}}{m^{2l+1}(2l+1)!} U \left( \overleftarrow{\nabla}_r \cdot \vec{\nabla}_v \right)^{2l} \left( \overleftarrow{\nabla}_r \times \vec{\nabla}_v \right) f^{1,2}. \quad (2.4)$$

Thirdly, from a physical point of view, the first term of approximation series (2.2) has a clear interpretation as an analogue of Newton's second law:

$$\langle \dot{v}_\mu \rangle_{1,2} = -\frac{1}{m}\frac{\partial U}{\partial x_\mu} + \frac{1}{f^{1,2}}\frac{\hbar^2}{4!m^3}\frac{\partial^3 U}{\partial x_\mu \partial x_\lambda \partial x_\nu}\frac{\partial^2 f^{1,2}}{\partial v_\lambda \partial v_\nu} - ...., \quad (2.5)$$

In the classical limit ($\hbar \to 0$) when the derivatives of functions $U$ and $f^{1,2}$ are bounded, approximation (2.5) becomes an analog of Newton's second law $m\langle \dot{\vec{v}} \rangle_{1,2} = -\nabla_r U$. A similar result is obtained according to Theorem 3 when (2.2) is averaged over the velocity space (2.3). Mean flow of accelerations $\langle \dot{v}_\mu \rangle_1$ corresponds to the external force in the equation of motion (i.6).

The transition from expression (1.1) to approximation (2.2) due to the Helmholtz expansion (1.2) is not unique. Let us consider analogs of approximation (2.2). We write expression (1.1), grouping the summands at the same powers $\hbar$, for example, for the first two powers, $\hbar^0$ and $\hbar^2$, we obtain:

at $\hbar^0$

$$U\left( \overleftarrow{\nabla}_r \cdot \vec{\nabla}_v \right) f^{1,2} = \frac{\partial U}{\partial x}\frac{\partial f^{1,2}}{\partial v_x} + \frac{\partial U}{\partial y}\frac{\partial f^{1,2}}{\partial v_y} + \frac{\partial U}{\partial z}\frac{\partial f^{1,2}}{\partial v_z}, \quad (2.6)$$

at $\hbar^2$

$$U\left( \overleftarrow{\nabla}_r \cdot \vec{\nabla}_v \right)^3 f^{1,2} = \frac{\partial^3 f^{1,2}}{\partial v_x^3}\frac{\partial^3 U}{\partial x^3} + \frac{\partial^3 f^{1,2}}{\partial v_y^3}\frac{\partial^3 U}{\partial y^3} + \frac{\partial^3 f^{1,2}}{\partial v_z^3}\frac{\partial^3 U}{\partial z^3} + \quad (2.7)$$

$$+3\frac{\partial^3 f^{1,2}}{\partial v_x \partial v_y^2}\frac{\partial^3 U}{\partial x \partial y^2}+3\frac{\partial^3 f^{1,2}}{\partial v_x \partial v_z^2}\frac{\partial^3 U}{\partial x \partial z^2}+2\frac{\partial^3 f^{1,2}}{\partial v_x \partial v_y \partial v_z}\frac{\partial^3 U}{\partial x \partial y \partial z}+$$

$$+3\frac{\partial^3 f^{1,2}}{\partial v_x^2 \partial v_y}\frac{\partial^3 U}{\partial x^2 \partial y}+3\frac{\partial^3 f^{1,2}}{\partial v_y \partial v_z^2}\frac{\partial^3 U}{\partial y \partial z^2}+2\frac{\partial^3 f^{1,2}}{\partial v_x \partial v_y \partial v_z}\frac{\partial^3 U}{\partial x \partial y \partial z}+$$

$$+3\frac{\partial^3 f^{1,2}}{\partial v_x^2 \partial v_z}\frac{\partial^3 U}{\partial x^2 \partial z}+3\frac{\partial^3 f^{1,2}}{\partial v_y^2 \partial v_z}\frac{\partial^3 U}{\partial y^2 \partial z}+2\frac{\partial^3 f^{1,2}}{\partial v_x \partial v_y \partial v_z}\frac{\partial^3 U}{\partial x \partial y \partial z}.$$

According to expression (1.1), each of the summands (2.6)-(2.7) corresponds to a fragment of the expression $\frac{\partial}{\partial v_\mu}\left[f^{1,2}\langle \dot{v}_\mu\rangle_{1,2}\right]$. Potential $U$ does not depend on velocity $\vec{v}$, therefore, for the expression (2.6), the correspondence seems natural «↦»:

$$\frac{\partial}{\partial v_\mu}\left[f^{1,2}\langle \dot{v}_\mu\rangle_{1,2}\right] \mapsto \frac{\partial U}{\partial x_\mu}\frac{\partial f^{1,2}}{\partial v_\mu} = \frac{\partial}{\partial v_\mu}\left(f^{1,2}\frac{\partial U}{\partial x_\mu}\right) \Rightarrow \langle \dot{v}_\mu\rangle_{1,2} \mapsto \frac{\partial U}{\partial x_\mu}. \qquad (2.8)$$

Correspondence (2.8) is present in approximation (2.2)/(2.5) and is natural from a physics perspective (see Remark 2). For summands (2.7) there are several options for constructing correspondences. Indeed, for the first three summands in expression (2.7), a correspondence of the following form seems to be logical:

$$f^{1,2}\langle \dot{v}_\mu\rangle_{1,2} \mapsto \frac{\partial^2 f^{1,2}}{\partial v_x^2}\frac{\partial^3 U}{\partial x^3}+\frac{\partial^2 f^{1,2}}{\partial v_y^2}\frac{\partial^3 U}{\partial y^3}+\frac{\partial^2 f^{1,2}}{\partial v_z^2}\frac{\partial^3 U}{\partial z^3}. \qquad (2.9)$$

For the remaining summands $\frac{\partial^3 f^{1,2}}{\partial v_\lambda^2 \partial v_\mu}$, $\frac{\partial^3 f^{1,2}}{\partial v_\lambda \partial v_\mu^2}$ and $\frac{\partial^3 f^{1,2}}{\partial v_\mu \partial v_\lambda \partial v_\upsilon}$ in expression (2.7), variants of the removal of velocity derivative $\frac{\partial}{\partial v_\mu}$ arise. Numerical coefficients for derivatives give an additional degree of freedom in making correspondences. Each subsequent group of summands with multiplier $\hbar^{2l}$ will contain more and more options for correspondence of derivatives. Thus, there is an infinite number of options for constructing an approximation for field $\langle \dot{v}_\mu\rangle_{1,2}$. Note that for polynomial potential (i.3) there are no mixed derivatives $\frac{\partial^n U}{\partial x_\mu \partial x_\lambda ...}$, which leads to certainty in choosing the correspondences (2.8)–(2.9).

From a physics perspective, only the first summand in (2.8) has a determination in the form of an analogue of Newton's second law, which defines the classical trajectory associated with the equation of motion (i.6). Subsequent summands in series (1.1) with multipliers $\hbar^{2l}$ introduce ambiguity/indeterminacy into the form of the expression for $\langle \dot{v}_\mu\rangle_{1,2}$. Formally, the representation for field $\langle \dot{v}_\mu\rangle_{1,2}$ can be written as Newton's second law:

$$m\langle \dot{v}_\mu\rangle_{1,2} = F_\mu^{1,2} = -\frac{\partial U}{\partial x_\mu}+O_\mu\left(\hbar^2\right), \qquad (2.10)$$

where summand $O_\mu(\hbar^2)$ can have various forms depending on combinations of summands like (2.7), and so on. All such combinations are of order in the Planck constant $\hbar^{2l}$, $l > 0$. If equation (2.10) is interpreted as an equation of motion, then summand $O_\mu(\hbar^2)$ can be considered as an indefinite force acting on a particle and deflecting it from the classical trajectory that satisfies the equation $m\langle \dot{\vec{v}} \rangle_{1,2} = -\nabla_r U$. Force $O_\mu(\hbar^2)$ introduces «indeterminacy» into the coordinates and velocities of the trajectory on scales $\hbar^{2l}$, $l > 0$.

It seems that in the classical approximation at $\hbar \to 0$, indeterminacy $O_\mu(\hbar^2)$ in equation (2.10) disappears, but this is not so. Let us consider in more detail the expression $O_\mu(\hbar^2)$ consisting of summands of the form:

$$O_\mu(\hbar^2) = \sum_{l=1}^{+\infty} \frac{(-1)^l}{4^l (2l+1)!} \frac{\hbar^{2l}}{W} \frac{\partial^{2l} W}{\partial p_\lambda ...} \frac{\partial^{2l+1} U}{\partial x_\nu ...} \qquad (2.11)$$

which is written with respect to the Wigner function. The derivative of the Wigner function (i.2) of order $2l$ in momentum variables has the form:

$$\underset{2l}{\frac{\partial^{2l} W}{\partial p_\lambda ...}} = \frac{(-1)^l}{\hbar^{2l}} \frac{1}{(2\pi\hbar)^3} \int_{\mathbb{R}^3} \langle ... \rangle e^{-i\frac{\vec{p}\cdot\vec{s}}{\hbar}} \underset{2l}{s_\lambda ...} d^3s, \qquad (2.12)$$

where components $s_\lambda$, $\lambda = 1...3$ of vector $\vec{s}$ correspond to coordinates $x_s, y_s, z_z$. Substituting (2.12) into representation (2.11) and separating the zero order of smallness in $\hbar$, we obtain

$$O_\mu(\hbar^2) = \sum_{l=1}^{+\infty} \frac{(-1)^l}{4^l (2l+1)!} \frac{\partial^{2l+1} U}{\partial x_\nu ...} \kappa + ..., \qquad (2.13)$$

where $\kappa$ is some function of coordinates and momentum with dimension $\left[ m^{2l} \right]$ (see (2.12), (i.2)). Thus, expression (2.13) for «corrective» force $O_\mu$ does not explicitly contain coefficients $\hbar^{2l}$, and the classical passage to the limit at $\hbar \to 0$, in the general case, will not lead to the disappearance of summand $O_\mu$. Depending on the type of potential $U$, the terms of series (2.13) will make a contribution comparable to $-\frac{\partial U}{\partial x_\mu}$ to the right-hand side of equation (2.10). For example, summands $U \sim r^{\pm n}$ ($n \in \mathbb{N}$, since $U$ is an analytic function) when substituting into series (2.13) are present in the form of summands $\frac{\partial^{2l+1} U}{\partial x_\nu ...}$, that is: $r^{n-3}, r^{n-5}, ...1$ or $r^{-n-3}, r^{-n-5}, ...$. On microscales $r < 1$, the main contribution to $O_\mu$ is given by summands $1, r$ or $r^{-n-3}, r^{-n-5}, ...$. Such summands are comparable to classical force $-\frac{\partial U}{\partial x_\mu}$ and will lead to a significant change in the trajectory on the micro-scale. On macroscales $r \gg 1$, the main contribution to the expression of $O_\mu$ is made by summands $r^{n-3}$, which are small compared to classical force $-\frac{\partial U}{\partial x_\mu} \sim r^{n-1}$.

Consequently, on macroscales, the trajectories defined by equation (2.10) are close to the classical ones.

## §3 Numerical results

As an example, let us consider a quantum system described by a probability density function [13, 14]:

$$f^1(\vec{r},t) = \frac{1}{(2\pi)^{3/2} \sigma_r^3} \exp\left[-\frac{r^2 \sin^2\theta + z^2}{2\sigma_r^2}\right], \tag{3.1}$$

$$\langle \vec{v} \rangle_1 (\vec{r}) = \frac{\hbar}{2m}\nabla_r \Phi = \frac{\hbar}{m}\nabla_r \varphi = \frac{\hbar}{m}\nabla_r \phi = \frac{\hbar}{m\rho}\vec{e}_\phi = \frac{\hbar}{mr\sin\theta}\vec{e}_\phi, \tag{3.2}$$

where $\rho = r\sin\theta$, coordinates $(\rho, \phi, z)$ correspond to the cylindrical coordinate system, and $(r, \theta, \phi)$ correspond to the spherical coordinate system. Functions (3.1), (3.2) are solutions of the first Vlasov equation (i.9) [9]. Function (3.1) satisfies the normalization condition $\int f^1 d^3 r = 1$. The first Vlasov equation (i.9) corresponds to the Schrödinger equation (3.3) with the Hamilton-Jacobi equation (3.4) [15, 12]:

$$i\hbar \frac{\partial}{\partial t}\Psi = -\frac{\hbar^2}{2m}\Delta_r \Psi + U\Psi, \quad f^1 = |\Psi|^2, \quad \arg\Psi = \varphi, \tag{3.3}$$

$$-\hbar \frac{\partial \varphi}{\partial t} = \frac{m}{2}|\langle \vec{v} \rangle|^2 + V, \quad V = U + Q, \quad Q = -\frac{\hbar^2}{2m}\frac{\Delta_r |\Psi|}{|\Psi|}, \tag{3.4}$$

where Q is the quantum potential [16, 17].

**Theorem 4** *Wave function*

$$\Psi(\rho, \phi, z, t) = \frac{1}{(2\pi)^{3/4} \sigma_r^{3/2}} \exp\left(-\frac{\rho^2 + z^2}{4\sigma_r^2} + i\phi - i\frac{E}{\hbar}t\right), \quad E = \frac{3\hbar^2}{4m\sigma_r^2}, \tag{3.5}$$

*is the solution of the Schrödinger equation (3.3) with the potential*

$$U(\rho, \phi, z) = \frac{\hbar^2}{8m\sigma_r^4}\left(\rho^2 + z^2 - 4\frac{\sigma_r^4}{\rho^2}\right), \tag{3.6}$$

*in this case, the quantum potential (3.4) has the form*

$$Q = \frac{\hbar^2}{4m\sigma_r^2}\left(3 - \frac{\rho^2 + z^2}{2\sigma_r^2}\right). \tag{3.7}$$

The proof of Theorem 4 is given in Appendix B.

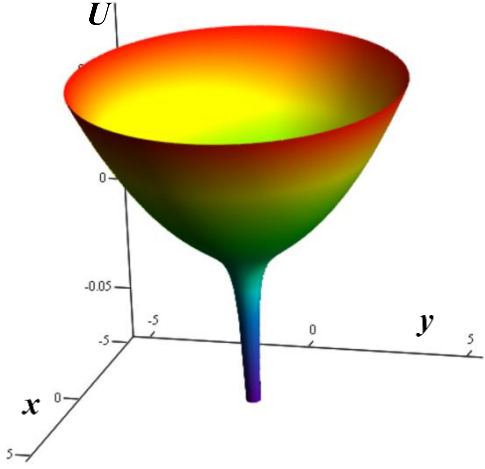

Fig. 1 Graph of potential $U(\rho,\phi,0)$

**Remark 4**

Note that the potential (3.6) on axis $OZ$ (at $\rho \to 0$) has a second-order pole, and at a distance from axis $OZ$ ($\rho \to +\infty$) it has the asymptotics of the harmonic oscillator potential (see Fig. 1). Fig. 1 shows the graph of potential $U$ at $z=0$. The pole of potential $U$ on axis $OZ$ arises due to vortex field $\langle \vec{v} \rangle_1$ (3.2). Scalar potential $\Phi$ is not a smooth function, therefore $\mathrm{curl}_r \langle \vec{v} \rangle_1 = \frac{\hbar}{2m} \mathrm{curl}_r \nabla_r \Phi \neq 0$. The properties of the extended form of the vortex field (3.2) are considered in detail in [14].

Potential $U$ having a pole leads to nonzero derivatives $\frac{\partial^{2l+1} U}{\partial x_\lambda ... \partial x_\mu}, l>0$ in the approximation of mean acceleration flux $\langle \dot{v}_\mu \rangle_{1,2}$ (2.5), (2.7). Consequently, additional force $O_\mu(\hbar^2)$ in equation of motion (2.10) will be different from zero.

**Remark 5**

The solution (3.5) of the Schrödinger equation corresponds to ground state of the quantum system $n=0$. Any function $f^1(\vec{r},t) = F(r)$ is solution $f^1$ of the first Vlasov equation (i.9) with vector field (3.2) [14]. Indeed,

$$\mathrm{div}_r \left[ f^1 \langle \vec{v} \rangle_1 \right] = F \mathrm{div}_r \langle \vec{v} \rangle_1 + \langle \vec{v} \rangle_1 \nabla_r F = \langle \vec{v} \rangle_1 \nabla_r F = \frac{\hbar}{m\rho} \frac{\partial F}{\partial r} \vec{e}_\phi \cdot \vec{e}_r = 0,$$

where $\nabla_r r = \vec{e}_r$. For example, $f_n^1(r) = c_n \cdot e^{-\frac{r^2}{2\sigma_r^2}} H_n^2(\eta r)$, where $c_n, \eta$ are constant values, and $H_n$ are Hermitian polynomials. Probability density functions $f_n^1(r)$ will correspond to the quantum states with numbers $n$.

The construction of approximation $\langle \dot{v}_\mu \rangle$ requires knowledge of the Wigner function for the quantum system under consideration. The following theorem is true.

**Theorem 5** *Wigner functiona $W$ and its derivativese $\frac{\partial^2 W}{\partial p_\lambda \partial p_\mu}$ corresponding to wave function $\Psi$ (3.5) can be represented as:*

$$G(\rho,z,\vec{p}) = \frac{\rho^2}{2\pi^4 \hbar^3 \sigma_r^2} e^{-\frac{\rho^2+z^2}{2\sigma_r^2} - \frac{2\sigma_r^2}{\hbar^2} p_z^2} \int_0^{2\pi} \frac{\Lambda(\rho,\bar{\rho}_s,\phi,\bar{\phi}_s,\vec{p})}{\sqrt{1+k_1^2 \sin^2 \bar{\phi}_s}} d\bar{\phi}_s \int_0^{+\infty} e^{-\rho^2 \frac{\bar{\rho}_s^2}{2\sigma_r^2}} \bar{\rho}_s d\bar{\rho}_s, \qquad (3.8)$$

*where*

$$\Lambda(\rho,\bar{\rho}_s,\phi,\bar{\phi}_s,\vec{p}) = \xi(\rho,\bar{\rho}_s,\phi+\bar{\phi}_s)(1+ik_1 \sin \bar{\phi}_s) e^{-i(p_\rho \cos \bar{\phi}_s + p_\phi \sin \bar{\phi}_s)\varsigma(\rho,\bar{\rho}_s)}, \qquad (3.9)$$

$$\xi(\rho,\bar{\rho}_s,\phi) = \begin{cases} 1, & \text{for } G = W, \\ -\zeta^2 \cos^2\phi, & \text{for } G = \dfrac{\partial^2 W}{\partial p_x^2}, \\ -\zeta^2 \sin^2\phi, & \text{for } G = \dfrac{\partial^2 W}{\partial p_y^2}, \\ -\dfrac{1}{2}\zeta^2 \sin 2\phi, & \text{for } G = \dfrac{\partial^2 W}{\partial p_x \partial p_y}, \end{cases} \qquad \zeta(\rho,\bar{\rho}_s) = 2\dfrac{\rho\bar{\rho}_s}{\hbar}, \qquad (3.10)$$

$$k_1(\bar{\rho}_s) = \frac{2\bar{\rho}_s}{1-\bar{\rho}_s^2}, \quad k_2(\bar{\rho}_s) = \frac{2\bar{\rho}_s}{1+\bar{\rho}_s^2}, \quad k_1^2 = \frac{k_2^2}{1-k_2^2}, \quad k_2^2 = \frac{k_1^2}{1+k_1^2}. \qquad (3.11)$$

The proof of Theorem 5 is given in Appendix B.

**Remark 6**

The integral (3.8) cannot be taken in quadratures, since even in a particular case of $p_\rho = p_\phi = 0$ it is expressed in terms of the complete elliptic integral of first kind $K(k)$. Indeed, (see Appendix C):

$$W(\vec{r}, p_z) = \frac{2\rho^2}{\pi^4 \hbar^3 \sigma_r^2} e^{-\frac{r^2}{2\sigma_r^2} - \frac{2\sigma_r^2}{\hbar^2} p_z^2} \int_0^{+\infty} K(ik_1) e^{-\frac{\rho^2 \bar{\rho}_s^2}{2\sigma_r^2}} \bar{\rho}_s d\bar{\rho}_s, \qquad (3.12)$$

where

$$K(k) = \int_0^{\pi/2} \frac{d\phi}{\sqrt{1-k^2 \sin^2\phi}}.$$

Integrand (3.8) can be rewritten in a compact form convenient for numerical calculations:

$$\frac{1+ik_1 \sin\bar{\phi}_s}{\sqrt{1+k_1^2 \sin^2\bar{\phi}_s}} e^{-i(p_\rho \cos\bar{\phi}_s + p_\phi \sin\bar{\phi}_s)\varsigma(\rho,\bar{\rho}_s)} = e^{i\theta(\rho,p_\rho,p_\phi,\bar{\rho}_s,\bar{\phi}_s)}, \qquad (3.13)$$

where

$$\theta(\rho, p_\rho, p_\phi, \bar{\rho}_s, \bar{\phi}_s) = \operatorname{arctg}(k_1 \sin\bar{\phi}_s) - (p_\rho \cos\bar{\phi}_s + p_\phi \sin\bar{\phi}_s)\varsigma(\rho,\bar{\rho}_s).$$

Note that averaging the momentum over the Wigner function (3.8)-(3.10) leads to the original expression for the velocity flux (3.2) (see Appendix C)

$$\langle \vec{p} \rangle_1 = m\langle \vec{v} \rangle_1 = \frac{1}{f^1(\rho,z)} \int_{\mathbb{R}^3} W(\rho,z,\vec{p}) \vec{p} d^3 p = \frac{\hbar}{\rho} \vec{e}_\phi. \qquad (3.14)$$

Although the Wigner function (3.8)–(3.10) is not known explicitly, the integral (3.14) can be calculated directly (see Appendix C).

The Wigner function (3.8)-(3.10) has azimuthal ($\phi$) symmetry and is defined in the 6D phase space. Fig. 2 shows distributions $W$ in the main phase planes. Fig. 2a shows the distribution in plane $(z, p_z)$ at $p_\rho = p_\phi = 0$ and $\rho = \sigma_r/2$. As follows from expression (3.12), in plane $(z, p_z)$ the Wigner function must have a Gaussian distribution, which is shown in Fig. 2a. Fig. 2b shows the distribution in plane $(\rho, p_\rho)$ at $z = 0$, $p_\phi = p_z = 0$, and Fig. 2c shows it in plane $(\rho, p_\phi)$ at $z=0$, $p_\rho = p_z = 0$. In accordance with the generalization of Hudson's theorem for 3D distributions of wave functions [4], the quasi-densities of the probabilities in Figs. 2b, 2c have areas of negative values.

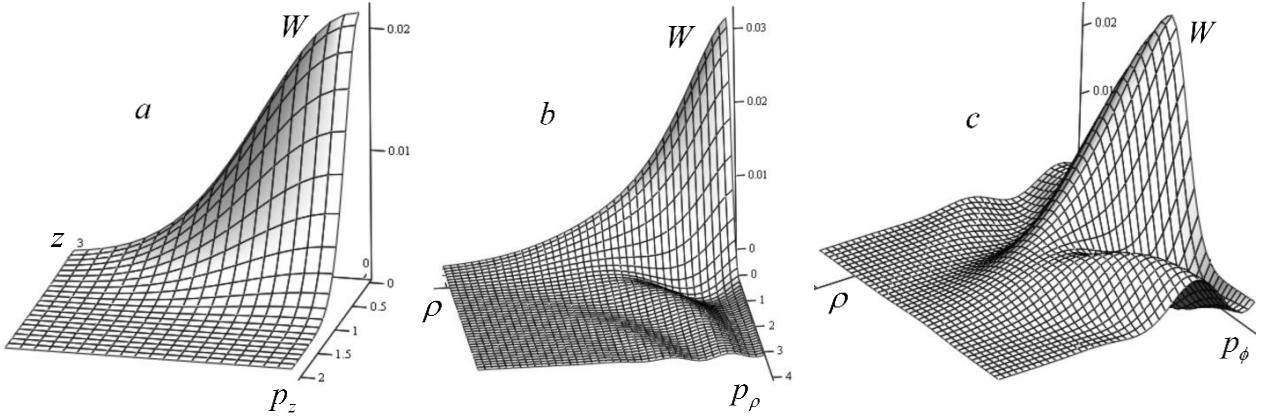

Fig. 2 Phase-plane portraits of Wigner function $W(\vec{r}, \vec{p})$

Expressions (3.2), (3.6) and (3.7) satisfy the Hamilton-Jacobi equation (3.4):

$$E = \frac{m}{2}|\langle \vec{v} \rangle|^2 + U + Q = const. \qquad (3.15)$$

Equation (3.15) defines a class of concentric circles (equations of motion) on the potential hypersurface (3.6) (see Fig. 1). Differentiating equation (3.15) with respect to the coordinate will give the equation of motion (i.6), in which $m\langle \dot{\vec{v}} \rangle_1 = -\nabla_r U$.

**Theorem 6** *Let the quantum system be described by the Wigner function (3.8)-(3.10), then the equation of motion (i.6) is valid, in which the pressure tensor (i.7) has the form:*

$$P_{\mu\lambda} = \frac{1}{m^2} \int_{(\infty)} W\left(p_\mu - \langle p_\mu \rangle_1\right)\left(p_\lambda - \langle p_\lambda \rangle_1\right) d^3p = f^1(\vec{r}) \frac{\hbar^2}{4m^2\sigma_r^2} \delta_{\mu\lambda}, \qquad (3.16)$$

$$\langle \dot{v}_\mu \rangle_1 = \frac{1}{f^1} \int_{(\infty)} W \langle \dot{v}_\mu \rangle_{1,2} d^3p = -\frac{1}{m} \frac{\partial U}{\partial x_\mu}, \qquad \frac{d}{dt} \langle \vec{v} \rangle_1 = -\frac{\hbar^2}{m^2 \rho^3} \vec{e}_\rho, \qquad (3.17)$$

*in this case, the quantum pressure force is related to the quantum potential as follows:*

$$\frac{\partial Q}{\partial x_\mu} = \frac{m}{f^1} \frac{\partial P_{\mu\lambda}}{\partial x_\lambda} = -\frac{\hbar^2 x_\mu}{4m\sigma_r^4}. \qquad (3.18)$$

The proof of Theorem 6 is given in Appendix B.

**Remark 7**

According to expression (3.2), the velocity of probability flow $\langle \vec{v} \rangle_1$ is directed along the tangents to the circles (3.15). Value (3.17) $\frac{d}{dt}\langle \vec{v} \rangle_1$ is the centripetal acceleration. Pressure tensor $P_{\mu\lambda}$ defines covariance matrix of momentum components $\text{cov}(p_\mu, p_\lambda)$, which has a diagonal form (3.16). It follows from the diagonal form of tensor $P_{\mu\lambda}$ that for the quantum system under consideration, the components of momentum $p_\mu$ are independent random variables.

To demonstrate the theoretical results presented in §2, we take the first two summands of expansion (2.6). The first summand in expansion (2.6)/(2.11), corresponding to the «classical» force, has the form:

$$\nabla_r U = \frac{\hbar^2}{4m\sigma_r^4}\left[\rho\left(1+4\frac{\sigma_r^4}{\rho^4}\right)\vec{e}_\rho + z\vec{e}_z\right]. \tag{3.19}$$

The second summand in the right-hand side of equation (2.11) can be estimated as

$$\frac{\partial^3 U}{\partial x_\nu \ldots} \sim \frac{\hbar^2}{m}\frac{1}{\rho^5}, \tag{3.20}$$

since

$$\frac{\partial^3 U}{\partial x^3} = \frac{12\hbar^2}{m\rho^5}\cos 2\phi \cos\phi, \quad \frac{\partial^3 U}{\partial y^3} = -\frac{12\hbar^2}{m\rho^5}\sin\phi\cos 2\phi,$$

$$\frac{\partial^3 U}{\partial x \partial y^2} = -\frac{4\hbar^2}{m\rho^5}\cos\phi(3\cos 2\phi - 2), \quad \frac{\partial^3 U}{\partial y \partial x^2} = \frac{4\hbar^2}{m\rho^5}\sin\phi(3\cos 2\phi + 2), \tag{3.21}$$

$$\frac{\partial^3 U}{\partial x \partial y \partial z} = \frac{\partial^3 U}{\partial x \partial z^2} = \frac{\partial^3 U}{\partial y \partial z^2} = \frac{\partial^3 U}{\partial z^3} = 0.$$

The expression for the Wigner function and its second-order derivatives is given in Theorem 5. Expressions (3.19) and (3.20) show that the characteristic spatial size of a physical system is value $\sigma_r$ equal to the standard deviation for the probability density (3.1). Value $\sigma_r$ is a free parameter in the solution (3.5)-(3.7). As an example, we consider a quantum system in which,

$$\sigma_r^2 = \frac{3}{2}r_0^2, \quad E = \frac{\hbar^2}{2mr_0^2}, \tag{3.22}$$

where $r_0$, $E$ are the radius and energy of the first Bohr orbit in a hydrogen atom, respectively. At $\rho < \sigma_r$ (microscale) the both contributions (3.19) and (3.20) have a significant impact on the trajectory, that satisfies equation of motion (2.11). On the macroscale at $\rho > \sigma_r$, the contribution (3.20) decreases, and equation (2.11) becomes the classical Newton equation.

The numerical calculations of the solutions to equation (2.11) will be carried out for three variations of approximation (2.6) of mean acceleration $\langle \dot{\vec{v}} \rangle$:

the 1st variation

$$U\left(\vec{\bar{\nabla}}_r \cdot \vec{\nabla}_v\right)^3 f^{1,2} \mapsto \begin{pmatrix} \dfrac{\partial^2 f^{1,2}}{\partial v_x^2} \dfrac{\partial^3 U}{\partial x^3} \\ \dfrac{\partial^2 f^{1,2}}{\partial v_y^2} \dfrac{\partial^3 U}{\partial y^3} \end{pmatrix} + 3 \begin{pmatrix} \dfrac{\partial^2 f^{1,2}}{\partial v_y^2} \dfrac{\partial^3 U}{\partial x \partial y^2} \\ \dfrac{\partial^2 f^{1,2}}{\partial v_x^2} \dfrac{\partial^3 U}{\partial x^2 \partial y} \end{pmatrix}, \qquad (3.23)$$

the 2nd variation

$$U\left(\vec{\bar{\nabla}}_r \cdot \vec{\nabla}_v\right)^3 f^{1,2} \mapsto \begin{pmatrix} \dfrac{\partial^2 f^{1,2}}{\partial v_x^2} \dfrac{\partial^3 U}{\partial x^3} \\ \dfrac{\partial^2 f^{1,2}}{\partial v_y^2} \dfrac{\partial^3 U}{\partial y^3} \end{pmatrix} + 3 \begin{pmatrix} \dfrac{\partial^2 f^{1,2}}{\partial v_x \partial v_y} \dfrac{\partial^3 U}{\partial x^2 \partial y} \\ \dfrac{\partial^2 f^{1,2}}{\partial v_x \partial v_y} \dfrac{\partial^3 U}{\partial x \partial y^2} \end{pmatrix}, \qquad (3.24)$$

the 3rd variation

$$\dfrac{\partial U}{\partial x_\mu}\left(\vec{\bar{\nabla}}_r \cdot \vec{\nabla}_v\right)^2 f^{1,2} \mapsto \begin{pmatrix} \dfrac{\partial^3 U}{\partial x^3} \dfrac{\partial^2 f^{1,2}}{\partial v_x^2} \\ \dfrac{\partial^3 U}{\partial y^3} \dfrac{\partial^2 f^{1,2}}{\partial v_y^2} \end{pmatrix} + \begin{pmatrix} \dfrac{\partial^3 U}{\partial x \partial y^2} \dfrac{\partial^2 f^{1,2}}{\partial v_y^2} \\ \dfrac{\partial^3 U}{\partial x^2 \partial y} \dfrac{\partial^2 f^{1,2}}{\partial v_x^2} \end{pmatrix} + 2 \dfrac{\partial^2 f^{1,2}}{\partial v_x \partial v_y} \begin{pmatrix} \dfrac{\partial^3 U}{\partial x^2 \partial y} \\ \dfrac{\partial^3 U}{\partial x \partial y^2} \end{pmatrix}. \qquad (3.25)$$

The third variation (3.25) corresponds to the Vlasov-Moyal approximation (2.2). Approximations (3.23)-(3.25) are written only for motion in plane $Z=0$, since the third-order derivatives of potential $U$ involving variable $z$ are equal to zero (3.21).

Fig. 3 (left) shows a classical trajectory with initial conditions $x = \sigma_r/2$, $y = z = 0$ and $p_x = p_z = 0$, $p_y = 3\sqrt{2mE}$. The arrows on the trajectory show the direction of particle motion. The coordinate scale $(\bar{x}, \bar{y})$ is in units of $\sigma_r$. On the right in Fig. 3 in addition to the classical trajectory (red) three trajectories (blue, green and black) are shown corresponding to approximations (3.23)-(3.25). On Fig. 3 (right) it can be seen that the trajectories have a significant difference between themselves, which begins to increase with time. The closest to the classical trajectory (red color) is the trajectory with the Vlasov-Moyal approximation (3.25) (black).

Note that the trajectories in Fig. 3 pass through the domains of micro- and macroscales. At the initial moment of time, the movement begins on the microscale ($\rho = \sigma_r/2$), where the difference in the approximations (3.23)-(3.25) is significant. This difference begins to show its influence almost at the very beginning as a splitting of trajectories (see Fig. 3 on the right). A similar situation is shown in Fig. 4 (left) for a smaller initial momentum $p_y = 1.641\sqrt{2mE}$. Fig. 4 illustrates the effect of the transition from the microscale to the macroscale. In Fig. 4 on the left, starting radius $\rho = \sigma_r/2$, center $\rho = 2.15\sigma_r$, and right $\rho = 2.5\sigma_r$. It can be seen that as the starting radius increases, all three trajectories (3.23)–(3.25) tend to the classical trajectory. Indeed, with a large starting radius (see Fig. 4 on the right), the influence of summands $O_\mu$ becomes small and the main contribution to the right-hand side of equation (2.11) is made by force $-\nabla_r U$ (classical trajectory). All four trajectories in Fig. 4 (right) are entirely within the macroscale domain. The plots of the trajectories in Fig. 4 (center) correspond to the transitional scale between micro- and macroscales. In Fig. 4 (left), there is a significant difference between the trajectories already at the initial moment of time, and in Fig. 4 (right) the difference between the trajectories appears only after a quarter of the period. In the central figure, a significant difference between the trajectories occurs at approximately one-eighth of the period.

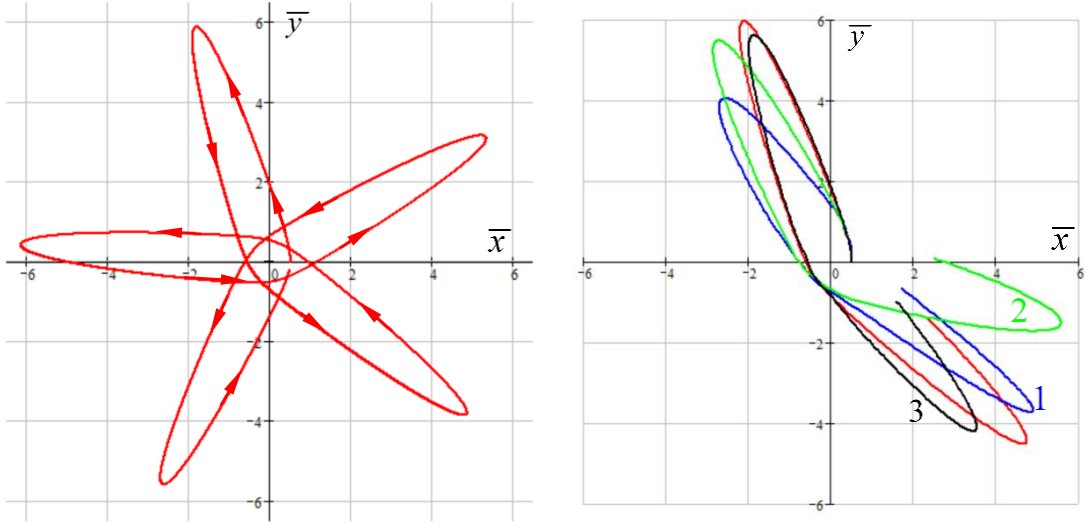

Fig. 3 Particle trajectories on the microscale

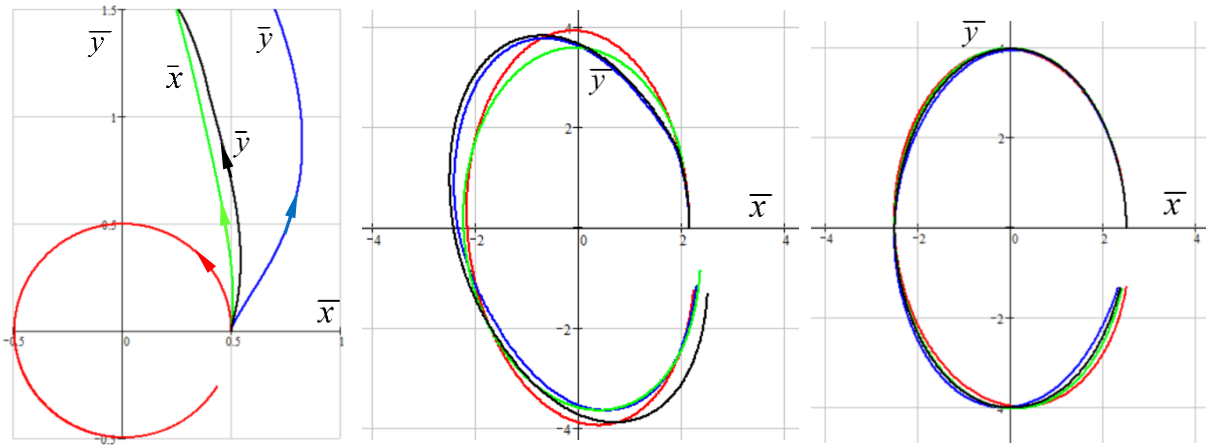

Fig. 4 Transition from the microscale to macroscale

All the above calculations are performed with the same value of the Planck constant ($\hbar = const$), that is, the classical transition ($\hbar \to 0$) is not performed anywhere. The contribution of additional force $O_\mu$ is determined only by the micro- and macroscales of the domain in which the trajectory is considered. The results of numerical simulation are in full agreement with the theoretical prediction made in §2.

**Remark 8**

The Wigner function has phase regions of negative and positive values (see Fig. 2), at the boundary of these regions, summand $O_\mu$ has poles (2.12). The presence of the poles leads to infinite energy barriers [18-21], near which particles can «reflect» sharply changing their trajectory.

The value of parameter $\sigma_r$ determines the spatial domain in which the particle trajectory is indeterminate (see Figs. 3, 4). In terms of probability density function $f^1 = |\Psi|^2$, parameter $\sigma_r$ is a root-mean-square deviation. The calculations presented in this section were performed at $\sigma_r \sim r_0$ (3.22). The size of the indeterminacy domain was of the order of the radius of the first Bohr orbit. In this case, the energy of the particle is small $E = \frac{3\hbar^2}{4m\sigma_r^2}$. Such a physical system on the microscale is close to a quantum system. If we assume that the minimum energy of a classical system equals to the rest mass energy $E = mc^2$, then

$$\sigma_r = \frac{\sqrt{3}}{2}\frac{\hbar}{mc} = \frac{\sqrt{3}}{2}\bar{\lambda}_C, \qquad (3.26)$$

where $\bar{\lambda}_C$ is the reduced Compton wavelength of the particle. Thus, the maximum indeterminacy of the center-of-mass of a classical particle will be proportional to its Compton wavelength.

**Conclusion**

Wave function $\Psi$ can be represented as a conformal mapping of some complex domain $\Psi = \exp(Z/2)$, where $Z = S + i\Phi$, and $S = \text{Ln}(\Psi\bar{\Psi})$ and $i\Phi = \text{Ln}(\Psi/\bar{\Psi})$ [12-13]. Function $S$ is related to the probability density $f^1$, and $\Phi$ is related to the scalar potential of the vector velocity field $\langle\vec{v}\rangle$ (3.2). There is an infinite set of trajectories along which points of one complex set, when mapping $\Psi$, can go to points of another complex set. For quantum mechanics, when mapping, density of «points» $f^1 = |\Psi|^2$ is important. The phase of wave function $\varphi = \Phi/2$ when mapping determines the mean vector field of probability flux $\langle\vec{v}\rangle_1$.

Vector field $\langle\vec{v}\rangle_1$ does not contain information about the motion of each point of the set, it is an averaging over the space of velocities (i.8). A similar picture is observed in hydro-gas dynamics, when the mean velocity of the medium is $\langle\vec{v}\rangle_1 = 0$, but each molecule/point is moving.

On the microscale in a gas/liquid it is impossible to determine the trajectory at point $\vec{r}_0$, for this it is necessary to average over a small volume around point $\vec{r}_0$ and obtain $\langle\vec{v}\rangle_1(\vec{r}_0)$ (i.8). A similar picture on the microscale is observed in Figs. 3 and 4 (left) when there is no complete information about acceleration field $\langle\dot{v}_\mu\rangle_{1,2}$. In a sense, for the Schrödinger equation, complete information about field $\langle\dot{v}_\mu\rangle_{1,2}$ is redundant, since quantum mechanics does not work with the trajectory, but with probability density $f^1$.

The Schrödinger equation can only give complete information about mean field $\langle\vec{v}\rangle_1 = \frac{\hbar}{2m}\nabla_r\Phi - \frac{e}{m}\vec{A}$, where $\vec{A}$ is the vector potential of magnetic field $\vec{B} = \text{curl}_r\vec{A}$ [15, 12]. A similar situation is observed on the macroscale in hydro-gas dynamics, in which one can operate with mean flow velocity of the medium $\langle\vec{v}\rangle_1$.

Thus, due to the mathematical formalism of Wigner-Vlasov, the similarity of interpretations of physical processes occurring in classical and quantum systems manifests itself.

**Acknowledgements**
This research has been supported by the Interdisciplinary Scientific and Educational School of Moscow University «Photonic and Quantum Technologies. Digital Medicine».

**Appendix**

*Proof of Theorem 1*

Substituting expression (1.1) into integral (1.3) and assuming that the term-by-term integration condition for series (1.1) is satisfied, we obtain:

$$\Pi(\vec{r},\vec{v},t) = \frac{1}{4\pi}\sum_{l=0}^{+\infty}\frac{(-1)^{l+1}(\hbar/2)^{2l}}{m^{2l+1}(2l+1)!}\int_{(\infty)}\frac{1}{|\vec{v}-\vec{v}'|}U\left(\vec{\nabla}_r\cdot\vec{\nabla}_v\right)^{2l+1}f^{1,2}d^3v' =$$

$$= \frac{1}{4\pi}\sum_{l=0}^{+\infty}\frac{(-1)^{l+1}(\hbar/2)^{2l}}{m^{2l+1}(2l+1)!}\int_{(\infty)}\frac{1}{|\vec{v}-\vec{v}'|}\sum_{k_1+k_2+k_3=2l+1}\frac{(2l+1)!}{k_1!k_2!k_3!}U\left(\frac{\bar{\partial}}{\partial x}\frac{\vec{\partial}}{\partial v'_x}\right)^{k_1}\left(\frac{\bar{\partial}}{\partial y}\frac{\vec{\partial}}{\partial v'_y}\right)^{k_3}\left(\frac{\bar{\partial}}{\partial z}\frac{\vec{\partial}}{\partial v'_z}\right)^{k_3}f^{1,2}d^3v' =$$

$$= \frac{1}{4\pi}\sum_{l=0}^{+\infty}\frac{(-1)^{l+1}(\hbar/2)^{2l}}{m^{2l+1}(2l+1)!}\sum_{k_1+k_2+k_3=2l+1}\frac{(2l+1)!}{k_1!k_2!k_3!}\frac{\partial^{2l+1}U}{\partial x^{k_1}\partial y^{k_2}\partial z^{k_3}}\times$$

$$\times\int_{-\infty}^{+\infty}\frac{1}{|\vec{v}-\vec{v}'|}\left(\frac{\vec{\partial}}{\partial v'_x}\right)^{k_1}f^{1,2}dv'_x\int_{-\infty}^{+\infty}\left(\frac{\vec{\partial}}{\partial v'_z}\right)^{k_3}dv'_z\int_{-\infty}^{+\infty}\left(\frac{\vec{\partial}}{\partial v'_y}\right)^{k_2}dv'_y,$$

(A.1)

Let us calculate separately the integral over $v'_x$:

$$\int_{-\infty}^{+\infty}\frac{1}{|\vec{v}-\vec{v}'|}\frac{\partial^{k_1}f^{1,2}}{\partial v'^{k_1}_x}dv'_x = \frac{1}{|\vec{v}-\vec{v}'|}\frac{\partial^{k_1-1}f^{1,2}}{\partial v'^{k_1-1}_x}\bigg|_{-\infty}^{+\infty} - \int_{-\infty}^{+\infty}\frac{\partial^{k_1-1}f^{1,2}}{\partial v'^{k_1-1}_x}\frac{\partial}{\partial v'_x}\frac{1}{|\vec{v}-\vec{v}'|}dv'_x = -\int_{-\infty}^{+\infty}\frac{\partial^{k_1-1}f^{1,2}}{\partial v'^{k_1-1}_x}\frac{\partial}{\partial v'_x}\frac{1}{|\vec{v}-\vec{v}'|}dv'_x =$$

$$= -\frac{\partial^{k_1-2}f^{1,2}}{\partial v'^{k_1-2}_x}\frac{\partial}{\partial v'_x}\frac{1}{|\vec{v}-\vec{v}'|}\bigg|_{-\infty}^{+\infty} + \int_{-\infty}^{+\infty}\frac{\partial^{k_1-2}f^{1,2}}{\partial v'^{k_1-2}_x}\frac{\partial^2}{\partial v'^2_x}\frac{1}{|\vec{v}-\vec{v}'|}dv'_x =$$

$$= \int_{-\infty}^{+\infty}\frac{\partial^{k_1-2}f^{1,2}}{\partial v'^{k_1-2}_x}\frac{\partial^2}{\partial v'^2_x}\frac{1}{|\vec{v}-\vec{v}'|}dv'_x = ... = (-1)^{k_1}\int_{-\infty}^{+\infty}f^{1,2}\frac{\partial^{k_1}}{\partial v'^{k_1}_x}\frac{1}{|\vec{v}-\vec{v}'|}dv'_x,$$

(A.2)

where the rapid decay to zero of the distribution function and its derivatives at $v\to\infty$ is taken into account. For the integrals over variables $v'_y$ and $v'_z$ one can obtain expressions similar to representation (A.2). Considering the above said and substituting (A.2) into (A.1), we obtain:

$$\Pi(\vec{r},\vec{v},t) = \frac{1}{4\pi}\sum_{l=0}^{+\infty}\frac{(-1)^{l+1}(\hbar/2)^{2l}}{m^{2l+1}(2l+1)!}\times$$

$$\sum_{k_1+k_2+k_3=2l+1}\frac{(2l+1)!}{k_1!k_2!k_3!}(-1)^{k_1}\frac{\partial^{2l+1}U}{\partial x^{k_1}\partial y^{k_2}\partial z^{k_3}}\int_{-\infty}^{+\infty}\frac{\partial^{k_1}}{\partial v'^{k_1}_x}\left(\frac{1}{|\vec{v}-\vec{v}'|}\right)\left(\frac{\vec{\partial}}{\partial v'_z}\right)^{k_3}\left(\frac{\vec{\partial}}{\partial v'_y}\right)^{k_2}f^{1,2}dv'_x\int_{-\infty}^{+\infty}dv'_z\int_{-\infty}^{+\infty}dv'_y =$$

$$\frac{1}{4\pi}\sum_{l=0}^{+\infty}\frac{(-1)^{l+1}(\hbar/2)^{2l}(-1)^{2l+1}}{m^{2l+1}(2l+1)!}\sum_{k_1+k_2+k_3=2l+1}\frac{(2l+1)!}{k_1!k_2!k_3!}\frac{\partial^{2l+1}U}{\partial x^{k_1}\partial y^{k_2}\partial z^{k_3}}\int_{(\infty)}f^{1,2}\frac{\partial^{2l+1}}{\partial v'^{k_1}_x\partial v'^{k_2}_y\partial v'^{k_3}_z}\frac{1}{|\vec{v}-\vec{v}'|}d^3v'=$$

$$=\frac{1}{4\pi}\sum_{l=0}^{+\infty}\frac{(-1)^l(\hbar/2)^{2l}}{m^{2l+1}(2l+1)!}\int_{(\infty)}f^{1,2}\sum_{k_1+k_2+k_3=2l+1}\frac{(2l+1)!}{k_1!k_2!k_3!}\frac{\partial^{2l+1}U}{\partial x^{k_1}\partial y^{k_2}\partial z^{k_3}}\frac{\partial^{2l+1}}{\partial v'^{k_1}_x\partial v'^{k_2}_y\partial v'^{k_3}_z}\frac{1}{|\vec{v}-\vec{v}'|}d^3v',$$

$$\Pi(\vec{r},\vec{v},t)=\frac{1}{4\pi}\sum_{l=0}^{+\infty}\frac{(-1)^l(\hbar/2)^{2l}}{m^{2l+1}(2l+1)!}U\int_{(\infty)}\left[\left(\vec{\nabla}_r\cdot\vec{\nabla}_{v'}\right)^{2l+1}\frac{1}{|\vec{v}-\vec{v}'|}\right]f^{1,2}d^3v'. \quad (A.3)$$

Note that

$$\frac{\partial}{\partial v'_\mu}\frac{1}{|\vec{v}-\vec{v}'|}=-\frac{\partial}{\partial v_\mu}\frac{1}{|\vec{v}-\vec{v}'|}\Rightarrow\frac{\partial}{\partial v'^k_\mu}\frac{1}{|\vec{v}-\vec{v}'|}=(-1)^k\frac{\partial}{\partial v_\mu}\frac{1}{|\vec{v}-\vec{v}'|}. \quad (A.4)$$

Considering transformation (A.4), expression (A.3) takes the form:

$$\Pi(\vec{r},\vec{v},t)=\frac{1}{4\pi}\sum_{l=0}^{+\infty}\frac{(-1)^l(\hbar/2)^{2l}}{m^{2l+1}(2l+1)!}\sum_{k_1+k_2+k_3=2l+1}\frac{(-1)^{2l+1}(2l+1)!}{k_1!k_2!k_3!}\frac{\partial^{2l+1}U}{\partial x^{k_1}\partial y^{k_2}\partial z^{k_3}}\frac{\partial^{2l+1}}{\partial v^{k_1}_x\partial v^{k_2}_y\partial v^{k_3}_z}\times$$

$$\times\int_{(\infty)}f^{1,2}\frac{1}{|\vec{v}-\vec{v}'|}d^3v'=\sum_{l=0}^{+\infty}\frac{(-1)^{l+1}(\hbar/2)^{2l}}{m^{2l+1}(2l+1)!}U\left(\vec{\nabla}_r\cdot\vec{\nabla}_v\right)^{2l+1}\vartheta,$$

where $\vartheta(\vec{r},\vec{v},t)=\dfrac{1}{4\pi}\displaystyle\int_{(\infty)}\frac{f^{1,2}(\vec{r},\vec{v}',t)}{|\vec{v}-\vec{v}'|}d^3v'$. Theorem 1 is proved.

*Proof of Theorem 2*
According to the potential theory, asymptotics $\vartheta^{1,2}$ has the form $\vartheta^{1,2}=C/|\vec{v}-\vec{v}^*|$ at $v\to\infty$, where $C=C(\vec{r},t)$ is some function, and $\vec{v}^*$ is some point from the space of velocities. Therefore, $\dfrac{\partial\vartheta^{1,2}}{\partial v_\mu}\sim\dfrac{1}{v^2}$ at $v\to\infty$. Potential (1.4) will have the asymptotics:

$$\Pi^{1,2}=-\frac{1}{m}U\left(\vec{\nabla}_r\cdot\vec{\nabla}_v\right)\vartheta^{1,2}+\sum_{l=1}^{+\infty}\frac{(-1)^{l+1}(\hbar/2)^{2l}}{m^{2l+1}(2l+1)!}U\left(\vec{\nabla}_r\cdot\vec{\nabla}_v\right)^{2l+1}\vartheta^{1,2}\sim\frac{1}{v^2}+O\left(\frac{\hbar^2}{v^4}\right),$$

from here

$$\frac{\partial}{\partial v_\mu}\Pi^{1,2}\sim\frac{1}{v^3}+O\left(\frac{\hbar^2}{v^5}\right). \quad (A.5)$$

It follows from expansion (1.2) that

$$\mathrm{div}_v\vec{J}^2_{1,2}=-\Delta_v\Pi^{1,2}, \quad (A.6)$$

$$\int_{\Sigma_{\infty}} \vec{J}_{1,2}^{2} \cdot d\vec{\sigma}_{v} = -\int_{\Sigma_{\infty}} \nabla_{v}\Pi^{1,2} \cdot d\vec{\sigma}_{v} =$$

$$= -\int_{-\infty}^{+\infty} dv_{y} \int_{-\infty}^{+\infty} dv_{z} \frac{\partial \Pi^{1,2}}{\partial v_{x}} \bigg|_{v_{x}=-\infty}^{v_{x}=+\infty} - \int_{-\infty}^{+\infty} dv_{x} \int_{-\infty}^{+\infty} dv_{x} \frac{\partial \Pi^{1,2}}{\partial v_{y}} \bigg|_{v_{y}=-\infty}^{v_{y}=+\infty} - \int_{-\infty}^{+\infty} dv_{x} \int_{-\infty}^{+\infty} dv_{y} \frac{\partial \Pi^{1,2}}{\partial v_{z}} \bigg|_{v_{z}=-\infty}^{v_{z}=+\infty} = 0, \quad (A.7)$$

where the asymptotics (A.5) is taken into account. A similar result is obtained when using distribution function $f^{1,2}$ that satisfies the Vlasov condition. For example, under the condition of term-by-term integration, for the first term of series (1.1), we obtain

$$\int_{(\infty)} \Delta_{v}\Pi^{1,2} d^{3}v = \frac{1}{m} U \int_{(\infty)} \left(\vec{\nabla}_{r} \cdot \vec{\nabla}_{v}\right) f^{1,2} d^{3}v + ... =$$

$$= \frac{1}{m}\frac{\partial U}{\partial x} \int_{(\infty)} \frac{\partial}{\partial v_{x}} f^{1,2} d^{3}v + \frac{1}{m}\frac{\partial U}{\partial y} \int_{(\infty)} \frac{\partial}{\partial v_{y}} f^{1,2} d^{3}v + \frac{1}{m}\frac{\partial U}{\partial z} \int_{(\infty)} \frac{\partial}{\partial v_{z}} f^{1,2} d^{3}v + .... = 0.$$

Expression (A.7) proves Theorem 2.

*Proof of Theorem 3*

Let us calculate $\text{div}_{v}\left[ f^{1,2} \left\langle \dot{\vec{v}} \right\rangle_{1,2} \right]$, we obtain

$$\frac{\partial}{\partial v_{\mu}}\left[ f^{1,2} \left\langle \dot{v}_{\mu} \right\rangle_{1,2} \right] = \sum_{l=0}^{+\infty} \frac{(-1)^{l+1}(\hbar/2)^{2l}}{m^{2l+1}(2l+1)!} \frac{\partial U}{\partial x_{\mu}} \left(\vec{\nabla}_{r} \cdot \vec{\nabla}_{v}\right)^{2l} \frac{\partial f^{1,2}}{\partial v_{\mu}} =$$

$$= \sum_{l=0}^{+\infty} \frac{(-1)^{l+1}(\hbar/2)^{2l}}{m^{2l+1}(2l+1)!} U \left(\vec{\nabla}_{r} \cdot \vec{\nabla}_{v}\right)^{2l+1} f^{1,2}. \quad (A.8)$$

Considering (A.8), the second Vlasov equation takes the form:

$$\partial_{t} f^{1,2} + \frac{\partial}{\partial v_{\mu}}\left[ f^{1,2} \left\langle \dot{v}_{\mu} \right\rangle_{1,2} \right] = 0, \quad (A.9)$$

$$\frac{\partial W}{\partial t} + \frac{1}{m}\vec{p}\cdot\nabla_{r}W - \nabla_{r}U\cdot\nabla_{p}W = \sum_{l=1}^{+\infty} \frac{(-1)^{l}(\hbar/2)^{2l}}{(2l+1)!} U\left(\vec{\nabla}_{r} \cdot \vec{\nabla}_{p}\right)^{2l+1} W = 0. \quad (A.10)$$

Let us integrate approximation (2.2) over the velocity space, we obtain

$$\int_{(\infty)} f^{1,2} \left\langle \dot{v}_{\mu} \right\rangle_{1,2} d^{3}v = f^{1} \left\langle \dot{v}_{\mu} \right\rangle_{1} = \frac{-1}{m}\frac{\partial U}{\partial x_{\mu}} \int_{(\infty)} f^{1,2} d^{3}v + \sum_{l=1}^{+\infty} \frac{(-1)^{l+1}(\hbar/2)^{2l}}{m^{2l+1}(2l+1)!} \frac{\partial U}{\partial x_{\mu}} \int_{(\infty)} \left(\vec{\nabla}_{r} \cdot \vec{\nabla}_{v}\right)^{2l} f^{1,2} d^{3}v =$$

$$= \frac{-1}{m}\frac{\partial U}{\partial x_{\mu}} f^{1} + \sum_{l=1}^{+\infty} \frac{(-1)^{l+1}(\hbar/2)^{2l}}{m^{2l+1}(2l+1)!} \frac{\partial U}{\partial x_{\mu}} \sum_{k_{1}+k_{2}+k_{3}=2l} \frac{\vec{\partial}^{2l}}{\partial x^{k_{1}}\partial y^{k_{2}}\partial z^{k_{3}}} \int_{(\infty)} \frac{\vec{\partial}^{2l} f^{1,2}}{\partial v_{x}^{k_{1}} \partial v_{y}^{k_{2}} \partial v_{z}^{k_{3}}} d^{3}v = \frac{-1}{m}\frac{\partial U}{\partial x_{\mu}} f^{1}, \quad (A.11)$$

where relations (i.8) and the Vlasov conditions at infinity for the distribution function are taken into account. Theorem 3 is proved.

*Proof of Theorem 4*

The form of the solution (3.5) follows directly from expressions (3.1) and (3.2) based on the approach described in [15]. The expression for potential $U$ can be obtained in two ways. The first way is to obtain it directly from the Schrödinger equation by substituting the wave function (3.5) into it. The second way is from the Hamilton-Jacobi equation (3.4). Without loss of generality, we use the first method:

$$\Psi_t = -i\frac{E}{\hbar}\Psi, \quad \Delta_r \Psi = \left(-\frac{3}{2\sigma_r^2} + \frac{\rho^2}{4\sigma_r^4} - \frac{1}{\rho^2} + \frac{z^2}{4\sigma_r^4}\right)\Psi. \tag{B.1}$$

Substituting expressions (B.1) into the Schrödinger equation, we obtain

$$U = E - \frac{3\hbar^2}{4m\sigma_r^2} + \frac{\hbar^2}{2m}\frac{\rho^2}{4\sigma_r^4} - \frac{\hbar^2}{2m}\frac{1}{\rho^2} + \frac{\hbar^2}{2m}\frac{z^2}{4\sigma_r^4}. \tag{B.2}$$

Taking into account the value of energy $E$ (3.5), expression (B.2) transforms into expression (3.6). To find quantum potential $Q$, we calculate $\Delta_r |\Psi|$:

$$|\Psi| = \frac{1}{(2\pi)^{3/4}\sigma_r^{3/2}}\exp\left(-\frac{\rho^2 + z^2}{4\sigma_r^2}\right), \quad \Delta_r |\Psi| = \left(\frac{\rho^2 + z^2}{4\sigma_r^4} - \frac{3}{2\sigma_r^2}\right)|\Psi|. \tag{B.3}$$

Expression (B.3) implies the validity of representation (3.7). Note that expressions (3.2), (3.6) and (3.7) satisfy the Hamilton-Jacobi equation (3.4):

$$E = \frac{m}{2}\frac{\hbar^2}{m^2\rho^2} + \frac{\hbar^2}{4m\sigma_r^2}\left(3 - \frac{\rho^2 + z^2}{2\sigma_r^2}\right) + \frac{\hbar^2}{8m\sigma_r^4}\left(\rho^2 + z^2 - 4\frac{\sigma_r^4}{\rho^2}\right). \tag{B.4}$$

Theorem 4 is completely proved.

*Proof of Theorem 5*

Let us obtain an expression for the Wigner function. We transform integrand (i.2), we obtain

$$\overline{\Psi}\left(\vec{r} - \frac{\vec{s}}{2}, t\right) = \frac{1}{(2\pi)^{3/4}\sigma_r^{3/2}}\exp\left(-\frac{1}{4\sigma_r^2}\left|\vec{r} - \frac{\vec{s}}{2}\right|^2 - i\phi_- + i\frac{E}{\hbar}t\right),$$

$$\Psi\left(\vec{r} + \frac{\vec{s}}{2}, t\right) = \frac{1}{(2\pi)^{3/4}\sigma_r^{3/2}}\exp\left(-\frac{1}{4\sigma_r^2}\left|\vec{r} + \frac{\vec{s}}{2}\right|^2 + i\phi_+ - i\frac{E}{\hbar}t\right),$$

$$\overline{\Psi}\left(\vec{r} - \frac{\vec{s}}{2}, t\right)\Psi\left(\vec{r} + \frac{\vec{s}}{2}, t\right) = \frac{e^{i(\phi_+ - \phi_-)}}{(2\pi)^{3/2}\sigma_r^3}e^{-\frac{1}{4\sigma_r^2}\left(\left|\vec{r} - \frac{\vec{s}}{2}\right|^2 + \left|\vec{r} + \frac{\vec{s}}{2}\right|^2\right)} = \frac{e^{i(\phi_+ - \phi_-)}}{(2\pi)^{3/2}\sigma_r^3}e^{-\frac{1}{2\sigma_r^2}\left(r^2 + \frac{s^2}{4}\right)}. \tag{B.5}$$

Azimuth angles $\phi_\pm$ in expression (B.5) satisfy the relations:

$$\sin\phi_\pm = \frac{2y \pm y_s}{\sqrt{(2x \pm x_s)^2 + (2y \pm y_s)^2}} = \frac{\sin\phi \pm \overline{\rho}_s \sin\phi_s}{\sqrt{1 + \overline{\rho}_s^2 \pm 2\overline{\rho}_s \cos(\phi - \phi_s)}}, \tag{B.6}$$

$$\cos\phi_{\pm} = \frac{2x \pm x_s}{\sqrt{(2x \pm x_s)^2 + (2y \pm y_s)^2}} = \frac{\cos\phi \pm \bar{\rho}_s \cos\phi_s}{\sqrt{1 + \bar{\rho}_s^2 \pm 2\bar{\rho}_s \cos(\phi - \phi_s)}},$$

where $\bar{\rho}_s = \dfrac{\rho_s}{2\rho}$ and it is taken into account that

$$\begin{aligned}(2x \pm x_s)^2 + (2y \pm y_s)^2 &= (2\rho\cos\phi \pm \rho_s\cos\phi_s)^2 + (2\rho\sin\phi \pm \rho_s\sin\phi_s)^2 = \\ &= 4\rho^2(\cos\phi \pm \bar{\rho}_s\cos\phi_s)^2 + 4\rho^2(\sin\phi \pm \bar{\rho}_s\sin\phi_s)^2 = \\ &= 4\rho^2\left[1 + \bar{\rho}_s^2 \pm 2\bar{\rho}_s(\cos\phi\cos\phi_s + \sin\phi\sin\phi_s)\right] = 4\rho^2\left[1 + \bar{\rho}_s^2 \pm 2\bar{\rho}_s\cos(\phi - \phi_s)\right].\end{aligned} \quad \text{(B.7)}$$

Using representation (B.6), we transform expression $e^{i(\phi_+ - \phi_-)}$:

$$e^{i(\phi_+ - \phi_-)} = \cos(\phi_+ - \phi_-) + i\sin(\phi_+ - \phi_-), \quad \text{(B.8)}$$

$$\cos(\phi_+ - \phi_-) = \cos\phi_+ \cos\phi_- + \sin\phi_+ \sin\phi_- =$$
$$= \frac{\cos^2\phi - \bar{\rho}_s^2 \cos^2\phi_s}{\sqrt{(1+\bar{\rho}_s^2)^2 - 4\bar{\rho}_s^2\cos^2(\phi-\phi_s)}} + \frac{\sin^2\phi - \bar{\rho}_s^2\sin^2\phi_s}{\sqrt{(1+\bar{\rho}_s^2)^2 - 4\bar{\rho}_s^2\cos^2(\phi-\phi_s)}},$$

$$\cos(\phi_+ - \phi_-) = \frac{1 - \bar{\rho}_s^2}{\sqrt{(1+\bar{\rho}_s^2)^2 - 4\bar{\rho}_s^2\cos^2(\phi-\phi_s)}}, \quad \text{(B.9)}$$

$$\sin(\phi_+ - \phi_-) = \sin\phi_+\cos\phi_- - \cos\phi_+\sin\phi_- =$$
$$= \frac{(\sin\phi + \bar{\rho}_s\sin\phi_s)(\cos\phi - \bar{\rho}_s\cos\phi_s) - (\cos\phi + \bar{\rho}_s\cos\phi_s)(\sin\phi - \bar{\rho}_s\sin\phi_s)}{\sqrt{(1+\bar{\rho}_s^2)^2 - 4\bar{\rho}_s^2\cos^2(\phi-\phi_s)}} =$$
$$= \frac{\sin\phi\cos\phi + \bar{\rho}_s\sin\phi_s\cos\phi - \bar{\rho}_s\sin\phi\cos\phi_s - \bar{\rho}_s^2\sin\phi_s\cos\phi_s}{\sqrt{(1+\bar{\rho}_s^2)^2 - 4\bar{\rho}_s^2\cos^2(\phi-\phi_s)}} -$$
$$- \frac{\cos\phi\sin\phi - \bar{\rho}_s\cos\phi\sin\phi_s + \bar{\rho}_s\cos\phi_s\sin\phi - \bar{\rho}_s^2\cos\phi_s\sin\phi_s}{\sqrt{(1+\bar{\rho}_s^2)^2 - 4\bar{\rho}_s^2\cos^2(\phi-\phi_s)}} =$$
$$= 2\bar{\rho}_s\frac{\sin\phi_s\cos\phi - \cos\phi_s\sin\phi}{\sqrt{(1+\bar{\rho}_s^2)^2 - 4\bar{\rho}_s^2\cos^2(\phi-\phi_s)}},$$
$$\sin(\phi_+ - \phi_-) = \frac{2\bar{\rho}_s\sin(\phi_s - \phi)}{\sqrt{(1+\bar{\rho}_s^2)^2 - 4\bar{\rho}_s^2\cos^2(\phi-\phi_s)}}. \quad \text{(B.10)}$$

We substitute expressions (B.9) and (B.10) into representation (B.8), we obtain

$$e^{i(\phi_+ - \phi_-)} = \frac{1 - \bar{\rho}_s^2 + i2\bar{\rho}_s\sin(\phi_s - \phi)}{\sqrt{(1+\bar{\rho}_s^2)^2 - 4\bar{\rho}_s^2\cos^2(\phi-\phi_s)}} = \frac{k_2}{k_1}\frac{1 + ik_1\sin(\phi_s - \phi)}{\sqrt{1 - k_2^2\cos^2(\phi-\phi_s)}}, \quad \text{(B.11)}$$

where $k_1 = \dfrac{2\bar{\rho}_s}{1-\bar{\rho}_s^2}$, $k_2 = \dfrac{2\bar{\rho}_s}{1+\bar{\rho}_s^2}$. Using a cylindrical coordinate system, we transform integrand $e^{-\frac{i}{\hbar}\vec{s}\cdot\vec{p}}$ into the Wigner functions:

$$\vec{p}\cdot\vec{s} = p_x x_s + p_y y_s + p_z z_s = p_x \rho_s \cos\phi_s + p_y \rho_s \sin\phi_s + p_z z_s, \tag{B.12}$$

$$p_x = p_\rho \cos\phi - p_\phi \sin\phi, \quad p_y = p_\rho \sin\phi + p_\phi \cos\phi. \tag{B.13}$$

We substitute (B.13) into (B.12), we obtain:

$$\vec{p}\cdot\vec{s} = \rho_s \left(p_\rho \cos\phi - p_\phi \sin\phi\right)\cos\phi_s + \rho_s \left(p_\rho \sin\phi + p_\phi \cos\phi\right)\sin\phi_s + p_z z_s =$$
$$= \rho_s p_\rho \left(\cos\phi\cos\phi_s + \sin\phi\sin\phi_s\right) + \rho_s p_\phi \left(\sin\phi_s \cos\phi - \cos\phi_s \sin\phi\right) + p_z z_s,$$
$$\vec{p}\cdot\vec{s} = \rho_s p_\rho \cos(\phi - \phi_s) + \rho_s p_\phi \sin(\phi_s - \phi) + p_z z_s. \tag{B.14}$$

Taking into account expressions (B.5), (B.11) and (B.14), the Wigner function will take the form:

$$W(\vec{r},\vec{p}) = \dfrac{1}{(2\pi\hbar)^3 (2\pi)^{3/2} \sigma_r^3} e^{-\frac{r^2}{2\sigma_r^2}} I(\rho,\vec{p}), \tag{B.15}$$

$$I(\rho,\vec{p}) = \int_{\mathbb{R}^3} \dfrac{k_2}{k_1} \dfrac{1+ik_1 \sin(\phi_s-\phi)}{\sqrt{1-k_2^2 \cos^2(\phi_s-\phi)}} e^{-\frac{s^2}{8\sigma_r^2}} e^{-\frac{i}{\hbar}\left(\rho_s p_\rho \cos(\phi_s-\phi)+\rho_s p_\phi \sin(\phi_s-\phi)+p_z z_s\right)} d^3s.$$

Note that integral $I(\rho,\vec{p})$ does not depend on $\phi$, since the integration is performed over the full period of trigonometric functions. Let us calculate integral $I(\rho,\vec{p})$ from expression (B.15).

$$I(\rho,\vec{p}) = \int_{-\infty}^{+\infty} e^{-\frac{z_s^2}{8\sigma_r^2}-\frac{i}{\hbar}p_z z_s} dz_s \int_0^{2\pi} \dfrac{1+ik_1 \sin\bar{\phi}_s}{\sqrt{1-k_2^2 \cos^2\bar{\phi}_s}} d\bar{\phi}_s \int_0^{+\infty} \dfrac{k_2}{k_1} e^{-\frac{\rho_s^2}{8\sigma_r^2}} e^{-\frac{i}{\hbar}\rho_s\left(p_\rho \cos\bar{\phi}_s + p_\phi \sin\bar{\phi}_s\right)} \rho_s d\rho_s. \tag{B.16}$$

The integral along axis $OZ$ can be found explicitly:

$$\int_{-\infty}^{+\infty} e^{-\frac{z_s^2}{8\sigma_r^2}-\frac{i}{\hbar}p_z z_s} dz_s = e^{-\frac{2\sigma_r^2}{\hbar^2}p_z^2} \int_{-\infty}^{+\infty} e^{-\left(\frac{z_s}{2\sqrt{2}\sigma_r}+\frac{i}{\hbar}\sqrt{2}\sigma_r p_z\right)^2} dz_s = 2\sqrt{2\pi}\sigma_r e^{-\frac{2\sigma_r^2}{\hbar^2}p_z^2}. \tag{B.17}$$

We transform the integrand in the second integral:

$$\dfrac{k_2}{k_1} \dfrac{1+ik_1 \sin\bar{\phi}_s}{\sqrt{1-k_2^2 \cos^2\bar{\phi}_s}} = \dfrac{k_2}{k_1}\dfrac{1+ik_1 \sin\bar{\phi}_s}{\sqrt{1-k_2^2+k_2^2 \sin^2\bar{\phi}_s}} = \dfrac{k_2}{k_1\sqrt{1-k_2^2}}\dfrac{1+ik_1 \sin\bar{\phi}_s}{\sqrt{1+k_1^2 \sin^2\bar{\phi}_s}} = \dfrac{1+ik_1 \sin\bar{\phi}_s}{\sqrt{1+k_1^2 \sin^2\bar{\phi}_s}}, \tag{B.18}$$

where it is taken into account that

$$\frac{k_2^2}{1-k_2^2} = \frac{4\bar{\rho}_s^2}{\left(1+\bar{\rho}_s^2\right)^2 - 4\bar{\rho}_s^2} = \frac{4\bar{\rho}_s^2}{1-2\bar{\rho}_s^2 + \bar{\rho}_s^4} = \frac{4\bar{\rho}_s^2}{\left(1-\bar{\rho}_s^2\right)^2} = k_1^2. \tag{B.19}$$

Substituting expressions (B.18) and (B.17) into integral (B.16), we obtain

$$I(\rho, \vec{p}) = 8\sqrt{2\pi}\sigma_r \rho^2 e^{-\frac{2\sigma_r^2}{\hbar^2}p_z^2} \int_0^{2\pi} \frac{1+ik_1 \sin\bar{\phi}_s}{\sqrt{1+k_1^2 \sin^2 \bar{\phi}_s}} d\bar{\phi}_s \int_0^{+\infty} e^{-\frac{\rho^2 \bar{\rho}_s^2}{2\sigma_r^2} - \frac{i}{\hbar}2\rho\bar{\rho}_s\left(p_\rho \cos\bar{\phi}_s + p_\phi \sin\bar{\phi}_s\right)} \bar{\rho}_s d\bar{\rho}_s. \tag{B.20}$$

where the substitution $2\rho\bar{\rho}_s = \rho_s$ is made, as $k_1 = k_1(\bar{\rho}_s)$. Substituting integral (B.20) into the expression for the Wigner function (B.20) gives expression (3.8)

$$W(\vec{r}, \vec{p}) = \frac{\rho^2}{2\pi^4 \hbar^3 \sigma_r^2} e^{-\frac{r^2}{2\sigma_r^2} - \frac{2\sigma_r^2}{\hbar^2}p_z^2} \int_0^{2\pi} \frac{1+ik_1 \sin\bar{\phi}_s}{\sqrt{1+k_1^2 \sin^2 \bar{\phi}_s}} d\bar{\phi}_s \int_0^{+\infty} e^{-\frac{\rho^2 \bar{\rho}_s^2}{2\sigma_r^2} - i\frac{2\rho}{\hbar}\bar{\rho}_s\left(p_\rho \cos\bar{\phi}_s + p_\phi \sin\bar{\phi}_s\right)} \bar{\rho}_s d\bar{\rho}_s, \tag{B.21}$$

It follows from expressions (B.13) that

$$p_\rho = p_x \cos\phi + p_y \sin\phi, \quad p_\phi = -p_x \sin\phi + p_y \cos\phi, \tag{B.22}$$

$$\frac{\partial}{\partial p_x} e^{-i\frac{2\rho}{\hbar}\bar{\rho}_s\left(p_\rho \cos\bar{\phi}_s + p_\phi \sin\bar{\phi}_s\right)} = -i\frac{2\rho}{\hbar}\bar{\rho}_s \left(\cos\phi\cos\bar{\phi}_s - \sin\phi\sin\bar{\phi}_s\right) e^{-i\frac{2\rho}{\hbar}\bar{\rho}_s\left(p_\rho \cos\bar{\phi}_s + p_\phi \sin\bar{\phi}_s\right)} =$$
$$= -i\frac{2\rho}{\hbar}\bar{\rho}_s \cos\left(\phi + \bar{\phi}_s\right) e^{-i\frac{2\rho}{\hbar}\bar{\rho}_s\left(p_\rho \cos\bar{\phi}_s + p_\phi \sin\bar{\phi}_s\right)} = -\frac{i}{\hbar}2\rho\bar{\rho}_s \cos\phi_s e^{-i\frac{2\rho}{\hbar}\bar{\rho}_s\left(p_\rho \cos\bar{\phi}_s + p_\phi \sin\bar{\phi}_s\right)}, \tag{B.23}$$

$$\frac{\partial^2}{\partial p_x^2} e^{-i\frac{2\rho}{\hbar}\bar{\rho}_s\left(p_\rho \cos\bar{\phi}_s + p_\phi \sin\bar{\phi}_s\right)} = -\frac{4\rho^2}{\hbar^2}\bar{\rho}_s^2 \cos^2\phi_s e^{-i\frac{2\rho}{\hbar}\bar{\rho}_s\left(p_\rho \cos\bar{\phi}_s + p_\phi \sin\bar{\phi}_s\right)}, \tag{B.24}$$

$$\frac{\partial}{\partial p_y} e^{-i\frac{2\rho}{\hbar}\bar{\rho}_s\left(p_\rho \cos\bar{\phi}_s + p_\phi \sin\bar{\phi}_s\right)} = -i\frac{2\rho}{\hbar}\bar{\rho}_s \left(\sin\phi\cos\bar{\phi}_s + \cos\phi\sin\bar{\phi}_s\right) e^{-i\frac{2\rho}{\hbar}\bar{\rho}_s\left(p_\rho \cos\bar{\phi}_s + p_\phi \sin\bar{\phi}_s\right)} =$$
$$= -i\frac{2\rho}{\hbar}\bar{\rho}_s \sin\left(\phi + \bar{\phi}_s\right) e^{-i\frac{2\rho}{\hbar}\bar{\rho}_s\left(p_\rho \cos\bar{\phi}_s + p_\phi \sin\bar{\phi}_s\right)} = -\frac{i}{\hbar}2\rho\bar{\rho}_s \sin\phi_s e^{-i\frac{2\rho}{\hbar}\bar{\rho}_s\left(p_\rho \cos\bar{\phi}_s + p_\phi \sin\bar{\phi}_s\right)}, \tag{B.25}$$

$$\frac{\partial^2}{\partial p_y^2} e^{-i\frac{2\rho}{\hbar}\bar{\rho}_s\left(p_\rho \cos\bar{\phi}_s + p_\phi \sin\bar{\phi}_s\right)} = -\frac{4\rho^2}{\hbar^2}\bar{\rho}_s^2 \sin^2\phi_s e^{-i\frac{2\rho}{\hbar}\bar{\rho}_s\left(p_\rho \cos\bar{\phi}_s + p_\phi \sin\bar{\phi}_s\right)}, \tag{B.26}$$

$$\frac{\partial^2}{\partial p_x \partial p_y} e^{-i\frac{2\rho}{\hbar}\bar{\rho}_s\left(p_\rho \cos\bar{\phi}_s + p_\phi \sin\bar{\phi}_s\right)} = -\frac{2\rho^2}{\hbar^2}\bar{\rho}_s^2 \sin 2\phi_s e^{-i\frac{2\rho}{\hbar}\bar{\rho}_s\left(p_\rho \cos\bar{\phi}_s + p_\phi \sin\bar{\phi}_s\right)}, \tag{B.27}$$

Taking into account expressions (B.24), (B.26) and (B.27) when differentiating the Wigner function $\frac{\partial^2 W}{\partial p_x^2}, \frac{\partial^2 W}{\partial p_y^2}, \frac{\partial^2 W}{\partial p_x \partial p_y}$, we obtain the validity of representations (3.10). Theorem 5 is proved.

*Proof of Theorem 6*

Let us transform the expression for pressure tensor $P_{\mu\lambda}$, we obtain:

$$m^2 P_{\mu\lambda} = \int\limits_{(\infty)} W\left(p_\mu - \langle p_\mu \rangle_1\right)\left(p_\lambda - \langle p_\lambda \rangle_1\right) d^3 p = \int\limits_{(\infty)} p_\mu p_\lambda W d^3 p - \langle p_\lambda \rangle_1 \int\limits_{(\infty)} p_\mu W d^3 p -$$
$$-\langle p_\mu \rangle_1 \int\limits_{(\infty)} p_\lambda W d^3 p + \langle p_\mu \rangle_1 \langle p_\lambda \rangle_1 \int\limits_{(\infty)} W d^3 p = \int\limits_{(\infty)} p_\mu p_\lambda W d^3 p - f^1 \langle p_\lambda \rangle_1 \langle p_\mu \rangle_1,$$
$$m^2 P_{\mu\lambda} = f^1 \left(\langle p_\mu p_\lambda \rangle_1 - \langle p_\lambda \rangle_1 \langle p_\mu \rangle_1\right). \tag{B.28}$$

Let us find an expression for $\langle p_x^2 \rangle$:

$$f^1 \langle p_x^2 \rangle = \frac{e^{-\frac{\rho^2+z^2}{2\sigma_r^2}}}{8\pi^4 \hbar^3 \sigma_r^2} \int\limits_{-\infty}^{+\infty} e^{-\frac{2\sigma_r^2}{\hbar^2} p_z^2} dp_z \int\limits_{-\infty}^{+\infty} p_x^2 e^{-\frac{i}{\hbar} p_x x_s} dp_x \int\limits_{-\infty}^{+\infty} \frac{1+ik_1 \sin\overline{\phi}_s}{\sqrt{1+k_1^2 \sin^2 \overline{\phi}_s}} dy_s \int\limits_{-\infty}^{+\infty} e^{-\frac{\rho_s^2}{8\sigma_r^2}} dx_s \int\limits_{-\infty}^{+\infty} e^{-\frac{i}{\hbar} p_y y_s} dp_y =$$

$$= \frac{\sqrt{2\pi} e^{-\frac{\rho^2+z^2}{2\sigma_r^2}}}{8\pi^3 \hbar \sigma_r^3} \int\limits_{-\infty}^{+\infty} p_x^2 e^{-\frac{i}{\hbar} p_x x_s} dp_x \int\limits_{-\infty}^{+\infty} \frac{1+ik_1 \sin\overline{\phi}_s}{\sqrt{1+k_1^2 \sin^2 \overline{\phi}_s}} e^{-\frac{y_s^2}{8\sigma_r^2}} \delta(y_s) dy_s \int\limits_{-\infty}^{+\infty} e^{-\frac{x_s^2}{8\sigma_r^2}} dx_s =$$

$$= -\frac{\hbar^2 \sqrt{2\pi} e^{-\frac{\rho^2+z^2}{2\sigma_r^2}}}{4\pi^2 \sigma_r^3} \int\limits_{-\infty}^{+\infty} \frac{1-ik_1(x_s)\sin\phi}{\sqrt{1+k_1^2(x_s)\sin^2\phi}} e^{-\frac{x_s^2}{8\sigma_r^2}} \delta''(x_s) dx_s,$$

$$f^1 \langle p_x^2 \rangle = -\frac{\hbar^2 \sqrt{2\pi} e^{-\frac{\rho^2+z^2}{2\sigma_r^2}}}{4\pi^2 \sigma_r^3} \frac{\partial^2}{\partial x_s^2} \frac{1-ik_1(x_s)\sin\phi}{\sqrt{1+k_1^2(x_s)\sin^2\phi}} e^{-\frac{x_s^2}{8\sigma_r^2}} \Bigg|_{x_s=0}, \tag{B.29}$$

where $\overline{\phi}_s = \phi_s - \phi = -\phi$ and it is taken into account that

$$\int\limits_{-\infty}^{+\infty} p_x^2 e^{-\frac{i}{\hbar} p_x x_s} dp_x = \hbar^3 \int\limits_{-\infty}^{+\infty} \overline{p}_x^2 e^{-i\overline{p}_x x_s} d\overline{p}_x = -2\pi\hbar^3 \delta''(x_s), \tag{B.30}$$

$$\int\limits_{-\infty}^{+\infty} F(x) \delta''(x-a) dx = F''(a). \tag{B.31}$$

Let us calculate the derivatives in expression (B.29):

$$\frac{\partial^2}{\partial x_s^2} \frac{1-ik_1(x_s)\sin\phi}{\sqrt{1+k_1^2(x_s)\sin^2\phi}} e^{-\frac{x_s^2}{8\sigma_r^2}} = \frac{\partial^2}{\partial x_s^2} \frac{1}{\sqrt{1+k_1^2 \sin^2\phi}} e^{-\frac{x_s^2}{8\sigma_r^2}} - i\frac{\partial^2}{\partial x_s^2} \frac{k_1 \sin\phi}{\sqrt{1+k_1^2 \sin^2\phi}} e^{-\frac{x_s^2}{8\sigma_r^2}}, \tag{B.32}$$

$$\frac{\partial}{\partial x_s} \frac{e^{-\frac{x_s^2}{8\sigma_r^2}}}{\sqrt{1+k_1^2 \sin^2\phi}} = -e^{-\frac{x_s^2}{8\sigma_r^2}} \frac{x_s \left(1+k_1^2 \sin^2\phi\right) + 4\sigma_r^2 k_1 k_1' \sin^2\phi}{4\sigma_r^2 \left(1+k_1^2 \sin^2\phi\right)^{3/2}},$$

$$\frac{\partial^2}{\partial x_s^2} \frac{e^{-\frac{x_s^2}{8\sigma_r^2}}}{\sqrt{1+k_1^2(x_s)\sin^2\phi}} = \frac{e^{-\frac{x_s^2}{8\sigma_r^2}}}{16\sigma_r^4} \frac{x_s^2 - 4\sigma_r^2}{\sqrt{1+k_1^2 \sin^2\phi}} -$$
$$- e^{-\frac{x_s^2}{8\sigma_r^2}} \sin^2\phi \frac{\left(k_1'^2 + k_1 k_1''\right)\left(1+k_1^2 \sin^2\phi\right) - 3k_1^2 k_1'^2 \sin^2\phi}{\left(1+k_1^2 \sin^2\phi\right)^{5/2}}. \tag{B.33}$$

Expression (B.33) at $x_s = 0$ takes the form:

$$\left.\frac{\partial^2}{\partial x_s^2} \frac{e^{-\frac{x_s^2}{8\sigma_r^2}}}{\sqrt{1+k_1^2(x_s)\sin^2\phi}}\right|_{x_s=0} = -\frac{1}{4\sigma_r^2} - k_1'^2 \sin^2\phi = -\frac{1}{4\sigma_r^2} - \frac{\sin^2\phi}{\rho^2}, \qquad (B.34)$$

where expression (C.10) for $k'$ is taken into account. We find the second summand from expression (B.32), we obtain

$$\frac{\partial}{\partial x_s} \frac{k_1 \sin\phi}{\sqrt{1+k_1^2\sin^2\phi}} e^{-\frac{x_s^2}{8\sigma_r^2}} = e^{-\frac{x_s^2}{8\sigma_r^2}} \sin\phi \left[\frac{4\sigma_r^2 k_1' - x_s k_1}{4\sigma_r^2 \sqrt{1+k_1^2\sin^2\phi}} - \frac{k_1^2 k_1' \sin^2\phi}{(1+k_1^2\sin^2\phi)^{3/2}}\right], \qquad (B.35)$$

$$\frac{\partial^2}{\partial x_s^2} \frac{k_1 \sin\phi}{\sqrt{1+k_1^2\sin^2\phi}} e^{-\frac{x_s^2}{8\sigma_r^2}} = -\frac{x_s}{4\sigma_r^2} \sin\phi\, e^{-\frac{x_s^2}{8\sigma_r^2}} \left[\frac{4\sigma_r^2 k_1' - x_s k_1}{4\sigma_r^2 \sqrt{1+k_1^2\sin^2\phi}} - \frac{k_1^2 k_1' \sin^2\phi}{(1+k_1^2\sin^2\phi)^{3/2}}\right] +$$

$$+ e^{-\frac{x_s^2}{8\sigma_r^2}} \sin\phi \frac{(4\sigma_r^2 k_1'' - k_1 - x_s k_1')(1+k_1^2\sin^2\phi) - (4\sigma_r^2 k_1' - x_s k_1)k_1 k_1'\sin^2\phi}{4\sigma_r^2 (1+k_1^2\sin^2\phi)^{3/2}} - \qquad (B.36)$$

$$- e^{-\frac{x_s^2}{8\sigma_r^2}} \sin^3\phi \frac{(2k_1 k_1'^2 + k_1^2 k_1'')(1+k_1^2\sin^2\phi)^{3/2} - 3k_1'^2 k_1^3 (1+k_1^2\sin^2\phi)^{1/2} \sin^2\phi}{(1+k_1^2\sin^2\phi)^3}.$$

Expression (B.36) at $x_s = 0$ takes the form:

$$\left.\frac{\partial^2}{\partial x_s^2} \frac{k_1 \sin\phi}{\sqrt{1+k_1^2\sin^2\phi}} e^{-\frac{x_s^2}{8\sigma_r^2}}\right|_{x_s=0} = \left.\frac{\partial^2 k_1}{\partial x_s^2}\right|_{x_s=0} \sin\phi = 0, \qquad (B.37)$$

where

$$\frac{\partial^2 k_1}{\partial x_s^2} = 4\rho \frac{\partial}{\partial x_s} \frac{4\rho^2 + x_s^2}{(4\rho^2 - x_s^2)^2} = 8\rho \frac{12\rho^2 + x_s^2}{(4\rho^2 - x_s^2)^3} x_s.$$

Substituting expressions (B.34) and (B.37) into (B.32) and then into (B.29), we obtain a representation for $\langle p_x^2 \rangle$:

$$f^1 \langle p_x^2 \rangle = \frac{\hbar^2 \sqrt{2\pi} e^{-\frac{\rho^2+z^2}{2\sigma_r^2}}}{4\pi^2 \sigma_r^3} \left(\frac{1}{4\sigma_r^2} + \frac{\sin^2\phi}{\rho^2}\right) \Rightarrow \langle p_x^2 \rangle = \frac{\hbar^2}{4\sigma_r^2}\left(1 + \frac{4\sigma_r^2}{\rho^2}\sin^2\phi\right). \qquad (B.38)$$

Considering (B.38) according to (B.28), we obtain an expression for $P_{xx}$:

$$P_{xx} = \frac{f^1}{m^2}\left(\langle p_x^2\rangle - \langle p_x\rangle^2\right) = \frac{f^1}{m^2}\left[\frac{\hbar^2}{4\sigma_r^2}\left(1 + \frac{4\sigma_r^2}{\rho^2}\sin^2\phi\right) - \frac{\hbar^2}{\rho^2}\sin^2\phi\right] = f^1\frac{\hbar^2}{4m^2\sigma_r^2}. \quad (B.39)$$

Let us perform similar calculations to find components $P_{yy}$ and $P_{zz}$. Let's calculate $\langle p_y^2\rangle$:

$$f^1\langle p_y^2\rangle = \frac{e^{-\frac{\rho^2+z^2}{2\sigma_r^2}}}{8\pi^4\hbar^3\sigma_r^2}\int_{-\infty}^{+\infty}e^{-\frac{2\sigma_r^2}{\hbar^2}p_z^2}dp_z\int_{-\infty}^{+\infty}e^{-\frac{i}{\hbar}p_x x_s}dp_x\int_{-\infty}^{+\infty}\frac{(1+ik_1\sin\overline{\phi}_s)}{\sqrt{1+k_1^2\sin^2\overline{\phi}_s}}dy_s\int_{-\infty}^{+\infty}e^{-\frac{\rho_s^2}{8\sigma_r^2}}dx_s\int_{-\infty}^{+\infty}p_y^2 e^{-\frac{i}{\hbar}p_y y_s}dp_y =$$

$$= \frac{\sqrt{2\pi}e^{-\frac{\rho^2+z^2}{2\sigma_r^2}}}{8\pi^3\hbar\sigma_r^3}\int_{-\infty}^{+\infty}p_y^2 e^{-\frac{i}{\hbar}p_y y_s}dp_y\int_{-\infty}^{+\infty}e^{-\frac{y_s^2}{8\sigma_r^2}}dy_s\int_{-\infty}^{+\infty}\frac{1+ik_1\sin\overline{\phi}_s}{\sqrt{1+k_1^2\sin^2\overline{\phi}_s}}\delta(x_s)e^{-\frac{x_s^2}{8\sigma_r^2}}dx_s =$$

$$= -\frac{\hbar^2\sqrt{2\pi}e^{-\frac{\rho^2+z^2}{2\sigma_r^2}}}{4\pi^2\sigma_r^3}\int_{-\infty}^{+\infty}\frac{1+ik_1(y_s)\cos\phi}{\sqrt{1+k_1^2(y_s)\cos^2\phi}}e^{-\frac{y_s^2}{8\sigma_r^2}}\delta''(y_s)dy_s,$$

$$f^1\langle p_y^2\rangle = -\frac{\hbar^2\sqrt{2\pi}e^{-\frac{\rho^2+z^2}{2\sigma_r^2}}}{4\pi^2\sigma_r^3}\frac{\partial^2}{\partial y_s^2}\frac{1+ik_1(y_s)\cos\phi}{\sqrt{1+k_1^2(x_s)\cos^2\phi}}e^{-\frac{y_s^2}{8\sigma_r^2}}\bigg|_{y_s=0}, \quad (B.40)$$

where $\overline{\phi}_s = \phi_s - \phi = \frac{\pi}{2} - \phi$ and $\int_{-\infty}^{+\infty}p_y^2 e^{-\frac{i}{\hbar}p_y y_s}dp_y = -2\pi\hbar^3\delta''(y_s)$ is taken into account. Let us find derivatives in expression (B.40). By analogy with transformations (B.32)-(B.37), we obtain:

$$\frac{\partial^2}{\partial y_s^2}\frac{1}{\sqrt{1+k_1^2(y_s)\cos^2\phi}}e^{-\frac{y_s^2}{8\sigma_r^2}}\bigg|_{y_s=0} = -\frac{1}{4\sigma_r^2} - \left(\frac{\partial k_1}{\partial y_s}\bigg|_{y_s=0}\right)^2\cos^2\phi = -\frac{1}{4\sigma_r^2} - \frac{\cos^2\phi}{\rho^2}, \quad (B.41)$$

$$\frac{\partial^2}{\partial y_s^2}\frac{k_1\cos\phi}{\sqrt{1+k_1^2\cos^2\phi}}e^{-\frac{y_s^2}{8\sigma_r^2}}\bigg|_{y_s=0} = \frac{\partial^2 k_1}{\partial y_s^2}\bigg|_{y_s=0}\cos\phi = 0. \quad (B.42)$$

Substituting derivatives (B.41) and (B.42) into expression (B.40), we find $\langle p_y^2\rangle$ and $P_{yy}$:

$$\langle p_y^2\rangle = \frac{\hbar^2}{4\sigma_r^2}\left(1 + \frac{4\sigma_r^2}{\rho^2}\cos^2\phi\right), \quad (B.43)$$

$$P_{yy} = \frac{f^1}{m^2}\left(\langle p_y^2\rangle - \langle p_y\rangle^2\right) = \frac{f^1}{m^2}\left(\frac{\hbar^2}{4\sigma_r^2}\left(1 + \frac{4\sigma_r^2}{\rho^2}\cos^2\phi\right) - \frac{\hbar^2}{\rho^2}\cos^2\phi\right) = f^1\frac{\hbar^2}{4m^2\sigma_r^2}. \quad (B.44)$$

Let us find an expression for $\langle p_z^2\rangle$:

$$f^1\langle p_z^2\rangle = \frac{e^{-\frac{\rho^2+z^2}{2\sigma_r^2}}}{8\pi^4\hbar^3\sigma_r^2}\int_{-\infty}^{+\infty}p_z^2 e^{-\frac{2\sigma_r^2}{\hbar^2}p_z^2}dp_z\int_{-\infty}^{+\infty}e^{-\frac{i}{\hbar}p_x x_s}dp_x\int_{-\infty}^{+\infty}\frac{1+ik_1\sin\overline{\phi}_s}{\sqrt{1+k_1^2\sin^2\overline{\phi}_s}}dy_s\int_{-\infty}^{+\infty}e^{-\frac{\rho_s^2}{8\sigma_r^2}}dx_s\int_{-\infty}^{+\infty}e^{-\frac{i}{\hbar}p_y y_s}dp_y =$$

$$= \sqrt{2\pi}\hbar^2 \frac{e^{-\frac{\rho^2+z^2}{2\sigma_r^2}}}{16\pi^2\sigma_r^5} \int_{-\infty}^{+\infty} \frac{(1+ik_1\sin\overline{\phi}_s)}{\sqrt{1+k_1^2\sin^2\overline{\phi}_s}} e^{-\frac{y_s^2}{8\sigma_r^2}} \delta(y_s) dy_s \int_{-\infty}^{+\infty} e^{-\frac{x_s^2}{8\sigma_r^2}} \delta(x_s) dx_s =$$

$$= \sqrt{2\pi}\hbar^2 \frac{e^{-\frac{\rho^2+z^2}{2\sigma_r^2}}}{16\pi^2\sigma_r^5} \int_{-\infty}^{+\infty} \frac{(1-ik_1(x_s)\sin\phi)}{\sqrt{1+k_1^2(x_s)\sin^2\phi}} e^{-\frac{x_s^2}{8\sigma_r^2}} \delta(x_s) dx_s,$$

$$f^1\langle p_z^2 \rangle = \frac{\sqrt{2\pi}\hbar^2}{16\pi^2\sigma_r^5} e^{-\frac{\rho^2+z^2}{2\sigma_r^2}}, \tag{B.45}$$

where it is taken into account that

$$\int_{-\infty}^{+\infty} p_z^2 e^{-\frac{2\sigma_r^2}{\hbar^2} p_z^2} dp_z = \sqrt{2\pi} \frac{\hbar^3}{8\sigma_r^3}.$$

From (B.45), we obtain expressions for $\langle p_z^2 \rangle$ and $P_{zz}$:

$$\langle p_z^2 \rangle = \frac{\hbar^2}{4\sigma_r^2}, \quad P_{zz} = \frac{1}{m^2} f^1\big(\langle p_z^2 \rangle - \langle p_z \rangle^2\big) = f^1 \frac{\hbar^2}{4m^2\sigma_r^2}, \tag{B.46}$$

where $\langle p_z \rangle = 0$ according to (3.2). Let us find mixed quantities $\langle p_\mu p_\lambda \rangle$, $\mu \neq \lambda$. Since $\langle p_z \rangle = 0$, then

$$\langle p_x p_z \rangle = \langle p_y p_z \rangle = 0,$$

from here

$$P_{xz} = \frac{1}{m^2} f^1\big(\langle p_x p_z \rangle - \langle p_x \rangle\langle p_z \rangle\big) = 0, \quad P_{yz} = \frac{1}{m^2} f^1\big(\langle p_y p_z \rangle - \langle p_y \rangle\langle p_z \rangle\big) = 0. \tag{B.47}$$

Thus, it remains to find $\langle p_x p_y \rangle$. Using the results (C.7) and (C.12), we obtain:

$$f^1\langle p_x p_y \rangle = \frac{e^{-\frac{\rho^2+z^2}{2\sigma_r^2}}}{8\pi^4\hbar^3\sigma_r^2} \int_{-\infty}^{+\infty} e^{-\frac{2\sigma_r^2}{\hbar^2}p_z^2} dp_z \int_{-\infty}^{+\infty} p_x e^{-\frac{i}{\hbar}p_x x_s} dp_x \int_{-\infty}^{+\infty} \frac{1+ik_1\sin\overline{\phi}_s}{\sqrt{1+k_1^2\sin^2\overline{\phi}_s}} dy_s \int_{-\infty}^{+\infty} e^{-\frac{\rho_s^2}{8\sigma_r^2}} dx_s \int_{-\infty}^{+\infty} p_y e^{-\frac{i}{\hbar}p_y y_s} dp_y =$$

$$= -\hbar^2\sqrt{2\pi} \frac{e^{-\frac{\rho^2+z^2}{2\sigma_r^2}}}{4\pi^2\sigma_r^3} \int_{-\infty}^{+\infty} \frac{1+ik_1\sin\overline{\phi}_s}{\sqrt{1+k_1^2\sin^2\overline{\phi}_s}} e^{-\frac{y_s^2}{8\sigma_r^2}} \delta'(y_s) dy_s \int_{-\infty}^{+\infty} \delta'(x_s) e^{-\frac{x_s^2}{8\sigma_r^2}} dx_s =$$

$$= -\hbar^2\sqrt{2\pi} \frac{e^{-\frac{\rho^2+z^2}{2\sigma_r^2}}}{4\pi^2\sigma_r^3} \int_{-\infty}^{+\infty} \frac{e^{-\frac{y_s^2}{8\sigma_r^2}}}{\sqrt{1+k_1^2\sin^2\overline{\phi}_s}} \delta'(y_s) dy_s \int_{-\infty}^{+\infty} \delta'(x_s) e^{-\frac{x_s^2}{8\sigma_r^2}} dx_s,$$

$$f^1\langle p_x p_y \rangle = -\hbar^2\sqrt{2\pi} \frac{e^{-\frac{\rho^2+z^2}{2\sigma_r^2}}}{4\pi^2\sigma_r^3} \frac{\partial^2}{\partial x_s \partial y_s} e^{-\frac{\rho_s^2}{8\sigma_r^2}} \frac{1}{\sqrt{1+k_1^2\sin^2\overline{\phi}_s}}\bigg|_{x_s=y_s=0}, \tag{B.48}$$

$$\int_{-\infty}^{+\infty} \delta'_y(y-b) dy \int_{-\infty}^{+\infty} F(x,y)\delta'_x(x-a) dx = -\int_{-\infty}^{+\infty} \delta'_y(y-b) \frac{\partial F}{\partial x}(a,y) dy = \frac{\partial^2 F}{\partial x \partial y}(a,b).$$

Let us transform the expression for derivative (B.48) at point $x_s = y_s = 0$ to the form:

$$\frac{\partial^2}{\partial x_s \partial y_s} e^{-\frac{\rho_s^2}{8\sigma_r^2}} \frac{1}{\sqrt{1+k_1^2 \sin^2 \bar{\phi}_s}}\bigg|_{x_s=y_s=0} = e^{-\frac{\rho_s^2}{8\sigma_r^2}} \frac{\partial^2}{\partial x_s \partial y_s} \frac{1}{\sqrt{1+k_1^2 \sin^2 \bar{\phi}_s}}\bigg|_{x_s=y_s=0} = \\ = \frac{\partial^2}{\partial x_s \partial y_s} \frac{1}{\sqrt{1+k_1^2 \sin^2 \bar{\phi}_s}}\bigg|_{x_s=y_s=0}. \quad (B.49)$$

Performing direct differentiation in expression (B.49), we obtain

$$\frac{\partial^2}{\partial x_s \partial y_s} e^{-\frac{\rho_s^2}{8\sigma_r^2}} \frac{1}{\sqrt{1+k_1^2 \sin^2 \bar{\phi}_s}}\bigg|_{x_s=y_s=0} = \frac{\sin\phi\cos\phi}{\rho^2}. \quad (B.50)$$

Substituting (B.50) into (B.48) gives expressions for $\langle p_x p_y \rangle$ and $P_{xy}$:

$$\langle p_x p_y \rangle = -\frac{\hbar^2}{\rho^2} \sin\phi\cos\phi, \quad (B.51)$$

$$P_{xy} = \frac{f^1}{m^2}(\langle p_x p_y \rangle - \langle p_x \rangle \langle p_y \rangle) = \frac{f^1}{m^2}\left(-\frac{\hbar^2}{\rho^2}\sin\phi\cos\phi + \frac{\hbar^2}{\rho^2}\sin\phi\cos\phi\right) = 0. \quad (B.52)$$

The resulting expressions (B.39), (B.43), (B.46), (B.47) and (B.52) prove the validity of assertion (3.16) of Theorem 6. Expression (3.17) for centripetal acceleration follows directly from differentiation of the stationary velocity field $\langle \vec{v} \rangle_1$ (3.2), that is $\frac{d}{dt}\langle v_\mu \rangle_1 = \langle v_\lambda \rangle_1 \frac{\partial}{\partial x_\lambda}\langle v_\mu \rangle_1$. The connection between quantum potential Q (3.5) and pressure tensor $P_{\mu\lambda}$ (3.18) is determined by the relation:

$$-\frac{m}{f^1}\frac{\partial P_{\mu\lambda}}{\partial x_\lambda} = -\frac{1}{mf^1}\frac{\hbar^2}{4\sigma_r^2}\delta_{\mu\lambda}\frac{\partial f^1}{\partial x_\lambda} = \frac{1}{mf^1}\frac{\hbar^2}{4\sigma_r^2}\delta_{\mu\lambda}\frac{x_\lambda}{\sigma_r^2}f^1 = \frac{\hbar^2 x_\mu}{4m\sigma_r^4} = \frac{\partial Q}{\partial x_\mu}, \quad (B.53)$$

where $\frac{\partial f^1}{\partial x_\lambda} = -\frac{x_\lambda}{\sigma_r^2}f^1$. Theorem 6 is completely proved.

**Appendix C**

The Wigner function (3.8) under the condition that $p_\rho = p_\phi = 0$ takes the form:

$$W(\vec{r}, p_z) = \frac{\rho^2}{2\pi^4 \hbar^3 \sigma_r^2} e^{-\frac{r^2}{2\sigma_r^2} - \frac{2\sigma_r^2}{\hbar^2}p_z^2} \int_0^{2\pi} \frac{1+ik_1 \sin\bar{\phi}_s}{\sqrt{1+k_1^2 \sin^2 \bar{\phi}_s}} d\bar{\phi}_s \int_0^{+\infty} e^{-\frac{\rho^2 \bar{\rho}_s^2}{2\sigma_r^2}} \bar{\rho}_s d\bar{\rho}_s. \quad (C.1)$$

Let us calculate the integral over angle $\bar{\phi}_s$, we obtain:

$$\int_0^{2\pi} \frac{1+ik_1 \sin \bar{\phi}_s}{\sqrt{1+k_1^2 \sin^2 \bar{\phi}_s}} d\bar{\phi}_s = \int_0^{2\pi} \frac{d\bar{\phi}_s}{\sqrt{1+k_1^2 \sin^2 \bar{\phi}_s}} + ik_1 \int_0^{2\pi} \frac{\sin \bar{\phi}_s}{\sqrt{1+k_1^2 \sin^2 \bar{\phi}_s}} d\bar{\phi}_s = 4K(ik_1), \quad \text{(C.2)}$$

where it is taken into account that the second integral equals zero. Indeed, according to (B.18)

$$\frac{ik_1 \sin \bar{\phi}_s}{\sqrt{1+k_1^2 \sin^2 \bar{\phi}_s}} = \frac{ik_2 \sin \bar{\phi}_s}{\sqrt{1-k_2^2 \cos^2 \bar{\phi}_s}}, \quad \text{(C.3)}$$

$$ik_1 \int_0^{2\pi} \frac{\sin \bar{\phi}_s}{\sqrt{1+k_1^2 \sin^2 \bar{\phi}_s}} d\bar{\phi}_s = i\int_0^{2\pi} \frac{k_2 \sin \bar{\phi}_s}{\sqrt{1-k_2^2 \cos^2 \bar{\phi}_s}} d\bar{\phi}_s = 0, \quad \text{(C.4)}$$

where $0 \leq k_2 < 1$ and the integrand has no poles.

Let us find mean momentum $\langle \vec{p} \rangle$ by averaging over the Wigner function. We rewrite the expression for the Wigner function in the Cartesian coordinate system, we obtain:

$$W(\rho, z, \vec{p}) = \frac{e^{-\frac{\rho^2+z^2}{2\sigma_r^2} - \frac{2\sigma_r^2}{\hbar^2} p_z^2}}{8\pi^4 \hbar^3 \sigma_r^2} \int_0^{2\pi} \frac{(1+ik_1 \sin \bar{\phi}_s) e^{-\frac{i}{\hbar}(p_\rho \rho_s \cos \bar{\phi}_s + p_\phi \rho_s \sin \bar{\phi}_s)}}{\sqrt{1+k_1^2 \sin^2 \bar{\phi}_s}} d\bar{\phi}_s \int_0^{+\infty} e^{-\frac{\rho_s^2}{8\sigma_r^2}} \rho_s d\rho_s =$$

$$= \frac{e^{-\frac{\rho^2+z^2}{2\sigma_r^2} - \frac{2\sigma_r^2}{\hbar^2} p_z^2}}{8\pi^4 \hbar^3 \sigma_r^2} \int_{-\infty}^{+\infty} \frac{(1+ik_1 \sin \bar{\phi}_s) e^{-\frac{i}{\hbar}(p_x x_s + p_y y_s)}}{\sqrt{1+k_1^2 \sin^2 \bar{\phi}_s}} dy_s \int_{-\infty}^{+\infty} e^{-\frac{\rho_s^2}{8\sigma_r^2}} dx_s, \quad \text{(C.5)}$$

where $\bar{\phi}_s = \phi_s - \phi$. Let us find $\langle p_x \rangle$:

$$f^1 \langle p_x \rangle = \frac{e^{-\frac{\rho^2+z^2}{2\sigma_r^2}}}{8\pi^4 \hbar^3 \sigma_r^2} \int_{-\infty}^{+\infty} e^{-\frac{2\sigma_r^2}{\hbar^2} p_z^2} dp_z \int_{-\infty}^{+\infty} p_x e^{-\frac{i}{\hbar} p_x x_s} dp_x \times$$

$$\times \int_{-\infty}^{+\infty} \frac{(1+ik_1 \sin \bar{\phi}_s)}{\sqrt{1+k_1^2 \sin^2 \bar{\phi}_s}} dy_s \int_{-\infty}^{+\infty} e^{-\frac{\rho_s^2}{8\sigma_r^2}} dx_s \int_{-\infty}^{+\infty} e^{-\frac{i}{\hbar} p_y y_s} dp_y. \quad \text{(C.6)}$$

Some integrals in expression (C.6) can be taken explicitly:

$$\int_{-\infty}^{+\infty} e^{-\frac{2\sigma_r^2}{\hbar^2} p_z^2} dp_z = \sqrt{2\pi} \frac{\hbar}{2\sigma_r}, \quad \int_{-\infty}^{+\infty} e^{-\frac{i}{\hbar} p_y y_s} dp_y = 2\pi \delta\left(\frac{y_s}{\hbar}\right) = 2\pi \hbar \delta(y_s), \quad \text{(C.7)}$$

$$\int_{-\infty}^{+\infty} p_x e^{-\frac{i}{\hbar} p_x x_s} dp_x = \hbar^2 \int_{-\infty}^{+\infty} \bar{p}_x e^{-i\bar{p}_x x_s} d\bar{p}_x = i\sqrt{2\pi}\sqrt{2\pi} \hbar^2 \delta'(x_s).$$

Substituting (C.7) into (C.6), we obtain

$$f^1\langle p_x\rangle = \frac{\sqrt{2\pi}e^{-\frac{\rho^2+z^2}{2\sigma_r^2}}}{8\pi^3\hbar\sigma_r^3}\int\limits_{-\infty}^{+\infty}p_x dp_x\int\limits_{-\infty}^{+\infty}\frac{\left(1+ik_1\sin\bar{\phi}_s\right)e^{-\frac{i}{\hbar}p_x x_s}}{\sqrt{1+k_1^2\sin^2\bar{\phi}_s}}\delta(y_s)dy_s\int\limits_{-\infty}^{+\infty}e^{-\frac{\rho_s^2}{8\sigma_r^2}}dx_s =$$

$$= \frac{\sqrt{2\pi}e^{-\frac{\rho^2+z^2}{2\sigma_r^2}}}{8\pi^3\hbar\sigma_r^3}\int\limits_{-\infty}^{+\infty}p_x e^{-\frac{i}{\hbar}p_x x_s}dp_x\int\limits_{-\infty}^{+\infty}\frac{1-ik_1(x_s)\sin\phi}{\sqrt{1+k_1^2(x_s)\sin^2\phi}}e^{-\frac{x_s^2}{8\sigma_r^2}}dx_s,$$

$$f^1\langle p_x\rangle = i\hbar\frac{\sqrt{2\pi}e^{-\frac{\rho^2+z^2}{2\sigma_r^2}}}{4\pi^2\sigma_r^3}\int\limits_{-\infty}^{+\infty}\delta'(x_s)\frac{1-ik_1(x_s)\sin\phi}{\sqrt{1+k_1^2(x_s)\sin^2\phi}}e^{-\frac{x_s^2}{8\sigma_r^2}}dx_s, \qquad (C.8)$$

where $k_1(x_s) = \frac{4\rho|x_s|}{4\rho^2 - x_s^2}$. Let us transform the remaining integral (C.9):

$$i\int\limits_{-\infty}^{+\infty}\delta'(x_s)\frac{1-ik_1\sin\phi}{\sqrt{1+k_1^2\sin^2\phi}}e^{-\frac{x_s^2}{8\sigma_r^2}}dx_s = i\int\limits_{-\infty}^{+\infty}\frac{\delta'(x_s)}{\sqrt{1+k_1^2\sin^2\phi}}e^{-\frac{x_s^2}{8\sigma_r^2}}dx_s + \sin\phi\int\limits_{-\infty}^{+\infty}\frac{\delta'(x_s)k_1}{\sqrt{1+k_1^2\sin^2\phi}}e^{-\frac{x_s^2}{8\sigma_r^2}}dx_s =$$

$$= -i\frac{\partial}{\partial x_s}\frac{e^{-\frac{x_s^2}{8\sigma_r^2}}}{\sqrt{1+k_1^2\sin^2\phi}}\bigg|_{x_s=0} - \sin\phi\frac{\partial}{\partial x_s}\frac{k_1 e^{-\frac{x_s^2}{8\sigma_r^2}}}{\sqrt{1+k_1^2\sin^2\phi}}\bigg|_{x_s=0} = -\frac{\sin\phi}{\rho}, \qquad (C.9)$$

where it is taken into account that

$$\int\limits_{-\infty}^{+\infty}F(x)\delta'(x-a)dx = -F'(a), \qquad (C.10)$$

and

$$\frac{\partial}{\partial x_s}e^{-\frac{x_s^2}{8\sigma_r^2}}\frac{1}{\sqrt{1+k_1^2\sin^2\phi}}\bigg|_{x_s=0} = -e^{-\frac{x_s^2}{8\sigma_r^2}}\frac{x_s}{4\sigma_r^2\sqrt{1+k_1^2\sin^2\phi}} - e^{-\frac{x_s^2}{8\sigma_r^2}}\frac{k_1 k_1'\sin^2\phi}{\left(1+k_1^2\sin^2\phi\right)^{3/2}}\bigg|_{x_s=0} = 0,$$

$$\frac{\partial}{\partial x_s}e^{-\frac{x_s^2}{8\sigma_r^2}}\frac{k_1}{\sqrt{1+k_1^2\sin^2\phi}}\bigg|_{x_s=0} = e^{-\frac{x_s^2}{8\sigma_r^2}}\frac{4\sigma_r^2 k_1' - x_s k_1}{4\sigma_r^2\sqrt{1+k_1^2\sin^2\phi}} - e^{-\frac{x_s^2}{8\sigma_r^2}}\frac{k_1 k_1'\sin^2\phi}{\left(1+k_1^2\sin^2\phi\right)^{3/2}}\bigg|_{x_s=0} = k_1' = \frac{1}{\rho}.$$

Substituting expressions (3.1) and (C.9) into (C.8), we obtain

$$f^1\langle p_x\rangle = \frac{e^{-\frac{\rho^2+z^2}{2\sigma_r^2}}}{(2\pi)^{3/2}\sigma_r^3}\langle p_x\rangle = -\hbar\frac{\sqrt{2\pi}e^{-\frac{\rho^2+z^2}{2\sigma_r^2}}}{4\pi^2\sigma_r^3}\frac{\sin\phi}{\rho},$$

$$\langle p_x\rangle = m\langle v_x\rangle = -2\pi\sqrt{2\pi}\hbar\frac{\sqrt{2\pi}}{4\pi^2}\frac{\sin\phi}{\rho} = -\hbar\frac{\sin\phi}{\rho},$$

$$\langle v_x\rangle = -\frac{\hbar}{m}\frac{\sin\phi}{\rho}. \qquad (C.11)$$

Similar calculations can be done for mean field $\langle p_y\rangle$.

$$f^1\langle p_y\rangle = \frac{e^{-\frac{\rho^2+z^2}{2\sigma_r^2}}}{8\pi^4\hbar^3\sigma_r^2}\int_{-\infty}^{+\infty}e^{-\frac{2\sigma_r^2}{\hbar^2}p_z^2}dp_z\int_{-\infty}^{+\infty}e^{-\frac{i}{\hbar}p_x x_s}dp_x\int_{-\infty}^{+\infty}\frac{(1+ik_1\sin\overline{\phi}_s)}{\sqrt{1+k_1^2\sin^2\overline{\phi}_s}}dy_s\int_{-\infty}^{+\infty}e^{-\frac{\rho_s^2}{8\sigma_r^2}}dx_s\int_{-\infty}^{+\infty}p_y e^{-\frac{i}{\hbar}p_y y_s}dp_y =$$

$$=\frac{\sqrt{2\pi}e^{-\frac{\rho^2+z^2}{2\sigma_r^2}}}{8\pi^3\hbar\sigma_r^3}\int_{-\infty}^{+\infty}\frac{(1+ik_1\sin\overline{\phi}_s)}{\sqrt{1+k_1^2\sin^2\overline{\phi}_s}}dy_s\int_{-\infty}^{+\infty}e^{-\frac{\rho_s^2}{8\sigma_r^2}}\delta(x_s)dx_s\int_{-\infty}^{+\infty}p_y e^{-\frac{i}{\hbar}p_y y_s}dp_y,$$

$$f^1\langle p_y\rangle = \frac{i\hbar\sqrt{2\pi}e^{-\frac{\rho^2+z^2}{2\sigma_r^2}}}{4\pi^2\sigma_r^3}\int_{-\infty}^{+\infty}e^{-\frac{y_s^2}{8\sigma_r^2}}\delta'(y_s)\frac{(1+ik_1(y_s)\cos\phi)}{\sqrt{1+k_1^2(y_s)\cos^2\phi}}dy_s, \qquad (C.12)$$

where

$$\int_{-\infty}^{+\infty}e^{-\frac{i}{\hbar}p_x x_s}dp_x = 2\pi\hbar\delta(x_s),\quad \int_{-\infty}^{+\infty}p_y e^{-\frac{i}{\hbar}p_y y_s}dp_y = i2\pi\hbar^2\delta'(y_s),\quad k_1(y_s)=\frac{4\rho|y_s|}{4\rho^2-y_s^2}.$$

Integral (C.12) is calculated similarly to integral (C.8) using formulas of (C.10):

$$i\int_{-\infty}^{+\infty}\delta'(y_s)e^{-\frac{y_s^2}{8\sigma_r^2}}\frac{(1+ik_1\cos\phi)}{\sqrt{1+k_1^2\cos^2\phi}}dy_s = i\int_{-\infty}^{+\infty}\frac{\delta'(y_s)}{\sqrt{1+k_1^2\cos^2\phi}}e^{-\frac{y_s^2}{8\sigma_r^2}}dy_s - \cos\phi\int_{-\infty}^{+\infty}\frac{\delta'(y_s)k_1}{\sqrt{1+k_1^2\cos^2\phi}}e^{-\frac{y_s^2}{8\sigma_r^2}}dy_s =$$

$$=-i\frac{\partial}{\partial y_s}\frac{e^{-\frac{y_s^2}{8\sigma_r^2}}}{\sqrt{1+k_1^2\cos^2\phi}}\bigg|_{y_s=0}+\cos\phi\frac{\partial}{\partial y_s}\frac{k_1 e^{-\frac{y_s^2}{8\sigma_r^2}}}{\sqrt{1+k_1^2\cos^2\phi}}\bigg|_{y_s=0}=\frac{\cos\phi}{\rho}, \qquad (C.13)$$

where it is taken into account that

$$\frac{\partial}{\partial y_s}e^{-\frac{y_s^2}{8\sigma_r^2}}\frac{1}{\sqrt{1+k_1^2\cos^2\phi}}\bigg|_{y_s=0}=-e^{-\frac{y_s^2}{8\sigma_r^2}}\frac{y_s}{4\sigma_r^2\sqrt{1+k_1^2\cos^2\phi}}-e^{-\frac{y_s^2}{8\sigma_r^2}}\frac{k_1 k_1'\cos^2\phi}{(1+k_1^2\cos^2\phi)^{3/2}}\bigg|_{y_s=0}=0,$$

$$\frac{\partial}{\partial y_s}e^{-\frac{y_s^2}{8\sigma_r^2}}\frac{k_1}{\sqrt{1+k_1^2\cos^2\phi}}\bigg|_{y_s=0}=e^{-\frac{y_s^2}{8\sigma_r^2}}\frac{4\sigma_r^2 k_1'-y_s k_1}{4\sigma_r^2\sqrt{1+k_1^2\cos^2\phi}}-e^{-\frac{y_s^2}{8\sigma_r^2}}\frac{k_1 k_1'\cos^2\phi}{(1+k_1^2\cos^2\phi)^{3/2}}\bigg|_{y_s=0}=k_1'=\frac{1}{\rho}.$$

Substituting expressions (3.1) and (C.13) into (C.12), we obtain:

$$\frac{e^{-\frac{\rho^2+z^2}{2\sigma_r^2}}}{(2\pi)^{3/2}\sigma_r^3}\langle p_y\rangle = \hbar\frac{\sqrt{2\pi}e^{-\frac{\rho^2+z^2}{2\sigma_r^2}}}{4\pi^2\sigma_r^3}\frac{\cos\phi}{\rho},$$

$$\langle v_y\rangle = \frac{\hbar}{m}\frac{\cos\phi}{\rho}. \qquad (C.14)$$

Since the Wigner function (C.5) is even in variable $p_z$, we obtain:

$$\langle v_z\rangle = 0. \qquad (C.15)$$

Passing to the cylindrical coordinate system (B.22) and taking into account expressions (C.11), (C.14) and (C.15) we arrive at expression (3.2):

$$\langle p_\rho \rangle = \langle p_x \rangle \cos\phi + \langle p_y \rangle \sin\phi = -\hbar \frac{\sin\phi}{\rho} \cos\phi + \hbar \frac{\cos\phi}{\rho} \sin\phi = 0,$$
$$\langle p_\phi \rangle = -\langle p_x \rangle \sin\phi + \langle p_y \rangle \cos\phi = \hbar \frac{\sin^2\phi}{\rho} + \hbar \frac{\cos^2\phi}{\rho} = \frac{\hbar}{\rho}.$$
(C.16)

20. Perepelkin E.E., Sadovnikov B.I., Inozemtseva N.G., Burlakov E.V., Afonin P.V., The Wigner function negative value domains and energy function poles of the polynomial oscillator, 2022, Physica A: Statistical Mechanics and its Applications, vol. 598, № 127339
21. Perepelkin E.E., Sadovnikov B.I., Inozemtseva N.G., Burlakov E.V., The Wigner function negative value domains and energy function poles of the harmonic oscillator, Journal of Computational Electronics, 2021, pp. 2148-2158
36